\documentclass[conference]{IEEEtran}
\IEEEoverridecommandlockouts
\usepackage{cite}
\usepackage{amsmath,amssymb,amsfonts}
\usepackage{algorithmic}
\usepackage{graphicx}
\usepackage{textcomp}
\usepackage{xcolor}

\usepackage[utf8]{inputenc} %
\usepackage[T1]{fontenc}    %
\usepackage{hyperref}       %
\usepackage{url}            %
\usepackage{booktabs}       %
\usepackage{amsfonts}       %
\usepackage{nicefrac}       %
\usepackage{microtype}      %

\usepackage{algorithmic}
\usepackage{graphicx}
\usepackage{textcomp}
\usepackage{xcolor}
\usepackage{multirow}
\usepackage{algorithm}
\usepackage{bm}
\usepackage{subfigure}
\usepackage{caption}
\usepackage{amsthm}

\def\BibTeX{{\rm B\kern-.05em{\sc i\kern-.025em b}\kern-.08em
    T\kern-.1667em\lower.7ex\hbox{E}\kern-.125emX}}
\begin{document}

\title{Memory Augmented Multi-Instance Contrastive Predictive Coding for Sequential Recommendation}

\author{\IEEEauthorblockN{Ruihong Qiu, Zi Huang, and Hongzhi Yin\IEEEauthorrefmark{1}\thanks{* Corresponding author.}}
\IEEEauthorblockA{\textit{School of Information Technology and Electrical Engineering, The University of Queensland,}\\
r.qiu@uq.edu.au, huang@itee.uq.edu.au, h.yin1@uq.edu.au}
}

\maketitle

\begin{abstract}\footnote{This work has been accepted by ICDM 2021.}
The sequential recommendation aims to recommend items, such as products, songs and places, to users based on the sequential patterns of their historical records. Most existing sequential recommender models consider the next item prediction task as the training signal. Unfortunately, there are two essential challenges for these methods: (1) the long-term preference is difficult to capture, and (2) the supervision signal is too sparse to effectively train a model. In this paper, we propose a novel sequential recommendation framework to overcome these challenges based on a memory augmented multi-instance contrastive predictive coding scheme, denoted as MMInfoRec. The basic contrastive predictive coding (CPC) serves as encoders of sequences and items. The memory module is designed to augment the auto-regressive prediction in CPC to enable a flexible and general representation of the encoded preference, which can improve the ability to capture the long-term preference. For effective training of the MMInfoRec model, a novel multi-instance noise contrastive estimation (MINCE) loss is proposed, using multiple positive samples, which offers effective exploitation of samples inside a mini-batch. The proposed MMInfoRec framework falls into the contrastive learning style, within which, however, a further finetuning step is not required given that its contrastive training task is well aligned with the target recommendation task. With extensive experiments on four benchmark datasets, MMInfoRec can outperform the state-of-the-art baselines.
\end{abstract}


\section{Introduction}
\label{sec:intro}
Recommender systems play an important role in today's online platform. To recommend proper items to a user, the essence is to predict the user's preference. However, the preference will naturally shift as time goes on, which requires the recommender systems to capture the dynamic preference of a user. Recently, various sequential models have been developed, which capture these preference dynamics, which are mainly based on sequence modeling methods~\cite{gru4rec,caser,sasrec,fdsa,bert4rec,s3rec,s2s,gru4recf,safm,fgnn,fgnnj,gag,posrec}. 

The most essential procedure of the sequential recommendation is to learn the sequential patterns of the user behaviors. Commonly, sequential recommender models rely on the next item for supervised training based on the assumption that if a model can predict exactly the user's preference, the model can successfully capture the sequential patterns~\cite{gru4rec,caser,sasrec}. The more recent methods introduce auxiliary tasks to enhance the training process. For example, some recent methods choose to use a manually masked content prediction task as the extra training signal, including masked item prediction~\cite{bert4rec} and masked attribute prediction~\cite{s3rec}. With the contrastive learning, the self-supervised learning over the sequence representation is applied by recent methods to enable the model to learn from different supervision signals~\cite{s3rec,s2s,xie2020contrastive,zhou2020contrastive}.

\begin{figure}
    \centering
    \includegraphics[width=\linewidth]{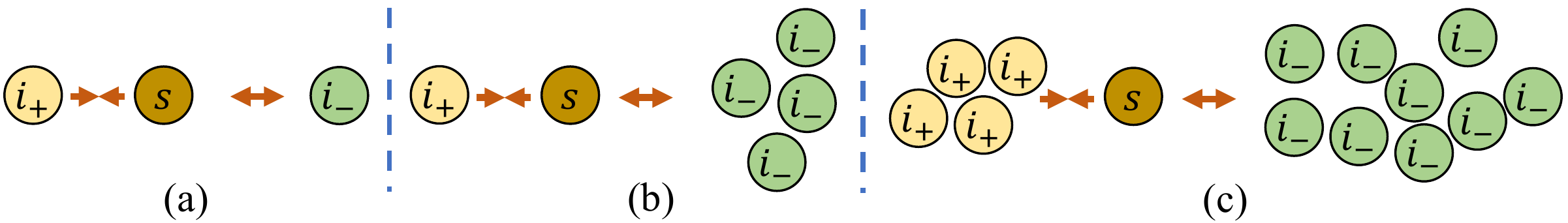}
    \caption{Illustration of latent representations of a sequence $s$ and its positive sample $i_+$ and its negative sample $i_-$ of different objectives. (a) A typical BPR objective with one representation of the next item serves as the positive sample and one negative sample. (b) Vanilla NCE objective with one ground truth positive sample and multiple negative samples. (c) Multi-instance of semantically positive and negative samples could alleviate the sparsity issue.}
    \label{fig:posneg}
\end{figure}

Although these methods have obtained a comparative progress in sequential recommendation, there are two essential challenges remained: (1) The long-term preference of a user is difficult to capture. The importance of the long-term preference is mainly indicated that the lone-term preference is related to the user's general interest, which can affect the user's behavior in addition to the user's most recent interest. Existing methods usually rely on RNN or Attention architectures to focus more on the latest interactions~\cite{gru4rec,sasrec}. Under these situations, the long-term preference fails to be emphasized. (2) The supervision signal is too sparse to effectively train a model. As shown in Fig.~\ref{fig:posneg} (a), the only objective directly related to the recommendation is still the next item prediction. Commonly, this objective is implemented by the cross-entropy loss~\cite{bert4rec} or Bayesian Personalized Ranking (BPR) loss~\cite{gru4rec}. In contrast, these proposed auxiliary tasks are all based on sequence understanding purposes and they are not aligned with the recommendation task. For example, it is not intuitive or straightforward how predicting a masked item, a masked short segment or a masked attribute in multiple previous steps can help predict the user's current preference.

Recently, there are two promising techniques that could solve the overcome the abovementioned challenges. Firstly, the external memory introduced in Neural Turing Machine~\cite{ntm} can explicitly store a set of memory slots for the long-term knowledge. The external memory has an effective reading and writing strategy for generating a comprehensive representation. Secondly, with the contrastive learning, the representation learning of different modals has achieved a great progress~\cite{moco,simclr,cpc,ecpc,deepinfomax,memdpc,coclr}. In these methods, optimizing a contrastive objective on different views of the same sample or the semantically similar samples can improve the quality of representation learning significantly by pulling the positive samples closer while pushing the negative samples further as in Fig.~\ref{fig:posneg} (b). For a sequence in the recommendation, it is desirable to pull input sequence and the predicted target item closer. In addition, when different views of the target item are incorporated into the contrastive learning, the number of semantically positive samples can be enlarged with a high quality, which is illustrated in Fig.~\ref{fig:posneg} (c).

In this paper, we aim to develop an effective solution to alleviate the sparsity issue of the sequential recommendation with the contrastive learning. Specifically, a memory augmented multi-instance contrastive predictive coding model is proposed, denoted as MMInfoRec. In the sequence encoding procedure, a contrastive predictive coding (CPC) scheme serves as a sequence encoder. A memory module is developed to augment the CPC in order to provide a flexible and general representation of the sequence representation to preserve the long-term preference effectively. For the design of the training objective, a multi-instance variant of the Noise Contrastive Estimation (NCE) is derived to provide a recommendation-related training signal to alleviate the sparse signal issue. The contributions of this paper are summarized as follows:
\begin{itemize}
    \item The MMInfoRec model is proposed for the sequential recommendation using an end-to-end contrastive learning scheme, which alleviates the sparse training signal issue in sequential recommendation.
    \item A memory module is developed to provide a flexible and general representation of the sequence, which can effectively preserve the long-term preference.
    \item A multi-instance contrastive loss, denoted as MINCE, is derived for the sequential recommendation task in the batch training setting, which provides rich training signals aligned with the recommendation.
    \item Extensive experiments on four datasets demonstrate the superiority of the MMInfoRec model compared with the state-of-the-art baselines.
\end{itemize}

This paper is organized as follows: In Section~\ref{sec:rw}, we will review the related literature. 
The detail of the proposed MMInfoRec is described in Section~\ref{sec:method}, followed by our extensive experiments in Section~\ref{sec:exp} to demonstrate the performance.

\section{Related Work}
\label{sec:rw}
\subsection{Sequential Recommendation}
\label{sec:sr}
With the successful usage of neural networks, many deep learning-based sequential recommender models have been developed~\cite{gru4rec,caser,sasrec,fdsa,bert4rec,s3rec,s2s,safm,zhou2020contrastive,xie2020contrastive,fgnn,fgnnj,gag,posrec,causalrec,lightweight,seq2graph}. They mainly use GRU~\cite{gru} or Transformer~\cite{attention} structures as the encoder. Some methods apply graph structure in sequential encoding such as FGNN~\cite{fgnn,fgnnj}, GAG~\cite{gag} and PosRec~\cite{posrec}. In terms of the training scheme, most of these methods follow the next-item supervised training style, for example, GRU4Rec~\cite{gru4rec}, Caser~\cite{caser}, SASRec~\cite{sasrec}, and FDSA~\cite{fdsa}. More recent methods apply masked token prediction by BERT4Rec~\cite{bert4rec}, masked segment prediction by ~$\text{S}^3$Rec\cite{s3rec}, attribute prediction by ~$\text{S}^3$Rec~\cite{s3rec}, or augmented sequences prediction by CL4SRec~\cite{xie2020contrastive} as contrastive training objective. BERT4Rec~\cite{bert4rec} and $\text{S}^3$Rec require a finetuning after the pre-training. Ma et al.~\cite{s2s} include the masked segment prediction as an extra objective as a component of multi-task learning along with the next-item prediction objective.

\subsection{Neural Memory Network}
\label{sec:nmn}
The neural memory network is a network module that mimics the mechanism of the memory in computers to support the writing and the reading for learning the long-term knowledge~\cite{ntm}. Recent recommendation models apply the memory module with different purposes~\cite{matn,magnn,nmrn,cmn,dman}. For example, NMRN~\cite{nmrn} makes use of the memory network to store the historical information in streaming recommendation to mitigate the catastrophic forgetting. Similarly, MA-GNN~\cite{magnn} and DMAN~\cite{dman} rely on the memory network to capture the long-term preference of the user. CMN~\cite{cmn} utilises the memory network to represent the user feature. MATN~\cite{matn} designs a memory network to learn the multiple behavior patterns.

\subsection{Contrastive Learning}
\label{sec:cl}
Contrastive learning has been used by recent methods for the sequential recommendation~\cite{s3rec,s2s,zhou2020contrastive,xie2020contrastive,channel,hypergraph}. Masked content prediction is applied by a few methods, including the masked item prediction and the masked attribute prediction by $\text{S}^3$Rec~\cite{s3rec} and the segment encoding by CL4SRec~\cite{xie2020contrastive}. These methods assume that all of the desired information is injected into the ID information through these pretext tasks.

Contrastive learning attracts interest in many areas, such as computer vision~\cite{moco,simclr,ecpc,deepinfomax,memdpc,ldsdg}, natural language processing~\cite{infoword} and speech representation learning~\cite{cpc}.
Recently, contrastive learning approaches achieve great success by discriminating the data itself among all other data~\cite{moco,simclr,ecpc}. These methods are based on estimating the mutual information between positive samples and negative samples. The mutual information is usually transformed into an optimizable bound~\cite{cpc,vbmi,deepinfomax}. For natural language processing and speech representation learning, pretext tasks include sequential prediction~\cite{cpc} and content prediction~\cite{infoword}. The sequential prediction objective requires the model to generate the following contents based on all the previous content.

\begin{figure*}
    \centering
    \includegraphics[width=\linewidth]{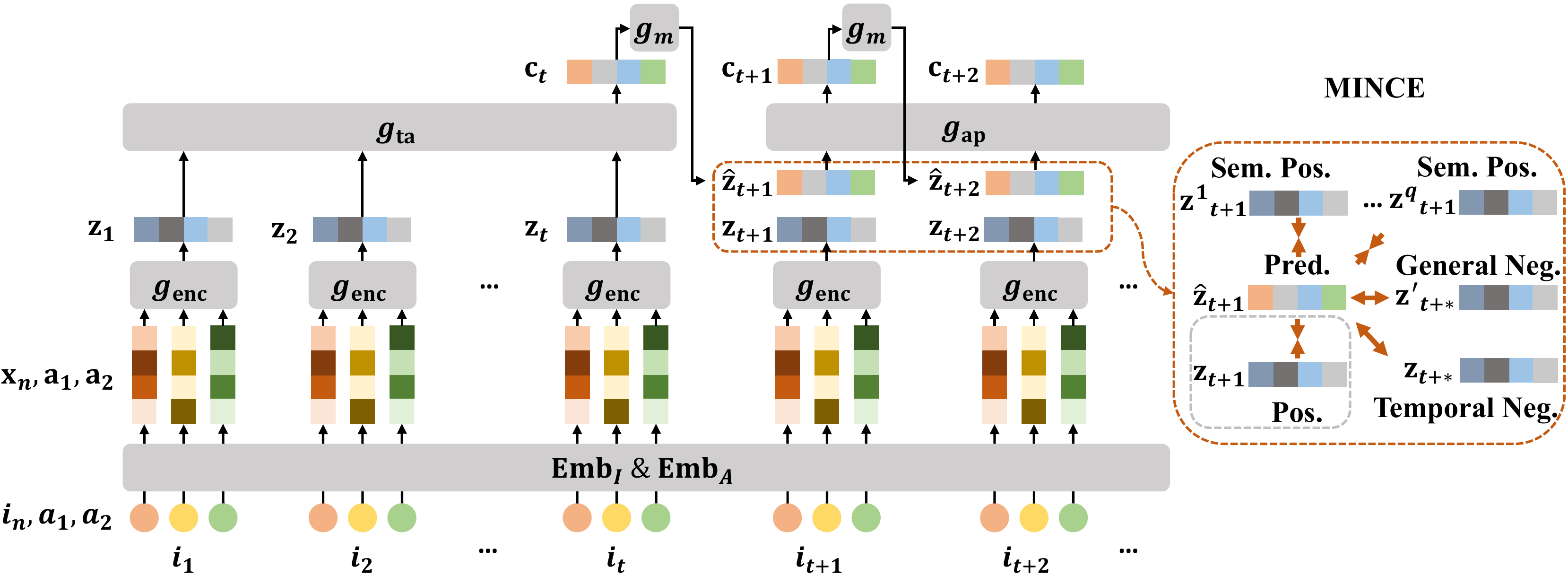}
    \caption{Overview of MMInfoRec. Every item is firstly encoded into a latent feature. For example, $i_t$ with its attributes in the sequence are encoded into a latent vector $\mathbf{z}_t$ by an encoder $g_\text{enc}$. The temporal aggregation module $g_\text{ta}$ can be any sequence encoding model such as GRU or Transformer. $\mathbf{c}_t$ is the result of the encoding of the sequence before time step $t$. In the prediction step $t+1$, a predicted result $\hat{\mathbf{z}}_{t+1}$ based on the memory function $g_\text{m}$ with $\mathbf{c}_t$ is compared to the ground truth latent vector $\mathbf{z}_{t+1}$, which is computed based on item $i_{t+1}$ and its attributes. The measurement of this comparison is MINCE, shown on the right for a two-step prediction. All of the positive samples (Pos.), the semantic positive samples (Sem.\ Pos.), the general negative samples (General Neg.) and the temporal negative samples (Temporal Neg.) are sampled from the training mini-batch. The semantic positive samples $\mathbf{z}^1_{t+1},\ldots,\mathbf{z}^q_{t+1}$ are calculated from different Dropout on the same $g_\text{enc}$. The general negative samples $\mathbf{z}'_{t+*}$ are from other sequences in the same batch and the temporal negative samples $\mathbf{z}_{t+*}$ are from the current sequence. With MINCE, the Pos.\ in the gray box will be omitted. If the Pos.\ is used instead of the Sem.\ Pos., then the MINCE degrades to NCE. During the test phase, the recommendation result is calculated based on the comparison among the latent representation $\mathbf{z}$.}
    \label{fig:whole}
\end{figure*}

\section{Method}
\label{sec:method}
Based on the abovementioned NCE loss, we will describe how to derive the MMInfoRec model in CPC scheme with the NCE loss in detail. In Section~\ref{sec:def}, we will provide the problem definition and mathematical notations. In Section~\ref{sec:frame}, the framework of the MMInfoRec will be described. The contrastive training objective is presented in Section~\ref{sec:con-loss}, followed by a discussion of the model in Section~\ref{sec:discuss}.

\subsection{Problem Definition and Notation}
\label{sec:def}
In sequential recommendation, there are usually a user set $\mathcal{U}$ and an item set $\mathcal{I}$, where $u\in \mathcal{U}$ denotes a user and $i\in \mathcal{I}$ denotes an item. $| \mathcal{U}|$ and $| \mathcal{I}|$ are used to represent the number of users and items respectively. Chronologically, a user $u$ has a historically ordered interaction sequence with multiple items: $\left[i_1,i_2,\ldots,i_n \right]$, where $n$ is the number of interactions. Besides, there is an attribute set $\mathcal{A}$ including all attributes belonging to all items. $| \mathcal{A}|$ is used to denote the number of attributes appearing in the dataset. For every item, there are associated attributes $A_i=\{a_1,a_2,\ldots,a_m\}$. Usually, different types of items have different attributes. For example, products have categories; songs are included in different albums; and videos are tagged with different genres. The purpose of a sequential recommender system is to predict the next item $i_{t+1}$ a user will interact with based on $\left[i_1,i_2,\ldots,i_t \right]$ and $A_i$ for each item, where $t$ indicates the current time step.

\subsection{Learning Framework}
\label{sec:frame}
The overview of the MMInfoRec is shown in Fig.~\ref{fig:whole}. The ID and attributes of all items $i$ in the sequence are firstly converted into dense feature vectors by two embedding layers respectively. For each item, its features are encoded by an attribute encoder $g_\text{enc}$ into latent representations $\mathbf{z}$. The temporal aggregation module $g_\text{ta}$ encodes the sequential information into $\mathbf{c}_t$, which is used to predict the futuristic latent representation by an auto-regressive prediction module $g_\text{ap}$.

\subsubsection{Embedding Layer}
\label{sec:emblay}
To convert IDs and attributes into dense vectors, two embedding matrices are applied: an item embedding matrix $\mathbf{Emb}_{I} \in \mathbb{R}^{|\mathcal{I}| \times d}$ and an attribute embedding matrix $\mathbf{Emb}_{A} \in \mathbb{R}^{|\mathcal{A}| \times d}$. $d$ represents the embedding size. For example, an item $i_n$ with its attributes $\{a_1,a_2\}$, we can use a lookup function to obtain the dense vectors: $\mathbf{x}_n=\mathbf{Emb}_{I}(i_n),\mathbf{a}_1=\mathbf{Emb}_{A}(a_1)$ and $\mathbf{a}_2=\mathbf{Emb}_{A}(a_2)$.

\subsubsection{Attribute Encoder}
\label{sec:attenc}
The attribute encoder aims to fuse all the information of the item, including the ID information and the side information, into a latent representation. After having the item embedding $\mathbf{x}$ and its attribute embeddings $\{\mathbf{a}_*\}$, an attribute encoder function $g_\text{enc}$ can compute the latent representation $\mathbf{z}$ of the item as:
\begin{equation}
    \mathbf{z}=g_\text{enc}(\mathbf{x},\{\mathbf{a}_*\}),
\end{equation}
where $\mathbf{x}\in\mathbb{R}^{1\times d}$, $\mathbf{a}_*\in\mathbb{R}^{1\times d}$ and $\mathbf{z}\in\mathbb{R}^{1\times d}$.

For the implementation of the function $g_\text{enc}$, a simple mean, or AutoInt~\cite{autoint}, or a self-attention of the item embedding to all attribute embeddings are all applicable for this structure.

\subsubsection{Temporal Aggregation}
\label{sec:temagg}
The temporal aggregation function is designed to aggregate the temporal information of the items before a certain time step. The function is defined as $g_\text{ta}$:
\begin{equation}
    \mathbf{c}_t=g_\text{ta}(\mathbf{z}_1,\mathbf{z}_1,\ldots,\mathbf{z}_t),
\end{equation}
where $\mathbf{c}_t\in\mathbb{R}^{1\times d}$.

For the choice of the temporal aggregation, common choices in sequential recommender models, such as GRU or Attention are capable of performing this aggregation computation.

\subsubsection{Memory Module}
To enhance the representation ability of the model, a memory module is designed to calculate the predictive output of the each step based on the context vector. In the memory module, there is a memory bank $\mathbf{M}\in\mathbb{R}^{b \times d}$ with $b$ memory slots. In the memory addressing is defined as:
\begin{align}
\label{eq:mem}
    \hat{\mathbf{z}}_{t+1}&=g_\text{m}\left(\mathbf{c}_{t}\right),\notag\\
    &=\operatorname{Softmax}\left(\operatorname{MLP}\left(\mathbf{c}_{t}\right)\right)\cdot\mathbf{M}+\mathbf{c}_{t},
\end{align}
where MLP stands for the multi layer perceptron. The residual style is designed for a retain of the original prediction and an improvement of the gradient flow in training.

\subsubsection{Auto-regressive Prediction}
For the multi-step prediction task, if the context is encoded into a vector $\mathbf{c}_t$, then the predicted latent representation $\mathbf{z}_{t+*}$ is expected to have strong semantic similarity with $\mathbf{c}_t$. Therefore, we introduce the auto-regressive prediction function $g_\text{ap}$,
\begin{equation}
\mathbf{c}_{t+1}=g_\text{ap}\left(\hat{\mathbf{z}}_{t+1}\right),\quad\hat{\mathbf{z}}_{t+2}=g_\text{m}\left(\mathbf{c}_{t+1}\right),
\end{equation}
where $\mathbf{c}_{t+1}$ is the summary of the context from time step $1$ to $t+1$, and $\hat{\mathbf{z}}_{t+2}$ is the predicted feature of the time step $t+2$. Similar to Seq2Seq style of prediction~\cite{seq2seq,zhou2020contrastive}, latent representations are predicted one-by-one in a sequential manner.

\subsubsection{Recommendation}
During the validation and test of MMInfoRec, the recommendation is conducted under a score ranking between the sequence representation $\mathbf{c}_t$ and all items in the item set. The score is simply computed with a dot-product between the sequence representation $\mathbf{c}_t$ at the current time step $t$ and the latent representations $\mathbf{z}$ of all items in the item set.

\subsection{Contrastive Loss}
\label{sec:con-loss}
Noise Contrastive Estimation (NCE)~\cite{nce} is a classification objective that can distinguish real samples and noisy samples. Similar to~\cite{nce2,cpc}, the NCE loss can be directly applied to the item predictive task. In MMInfoRec, we extend this loss to a multi-instance variant, which can effectively alleviate the issue of the sparse training signal.

\subsubsection{Vanilla NCE Loss}
The vanilla NCE mainly makes a comparison between representations in the latent space to force the predicted representation $\hat{\mathbf{z}}$ to be close to the ground truth representation $\mathbf{z}$. In the forward rollout, the MMInfoRec firstly computes the predicted latent vector $\hat{\mathbf{z}}$ and the ground truth representation $\mathbf{z}$. At time step $i$, the latent representations with the same size are denoted as $\hat{\mathbf{z}}_i$ and $\mathbf{z}_i$. The similarity of the predicted vector and ground-truth vector pair is calculated by the dot product $\mathbf{z}_i^\top\mathbf{z}_i$. The objective is:
\begin{equation}
\label{eq:loss}
    \ell_\text{NCE}=-\sum_{i}\left[\log \frac{e^{\left(\hat{\mathbf{z}}_{i}^{\top} \cdot \mathbf{z}_{i}\right)/\tau}}{e^{\left(\hat{\mathbf{z}}_{i}^{\top} \cdot \mathbf{z}_{i}\right)/\tau}+\sum\limits_{* \in \mathcal{N}_i} e^{\left(\hat{\mathbf{z}}_{i}^{\top} \cdot \mathbf{z}_{*}\right)/\tau}}\right],
\end{equation}
where $\tau$ is the temperature parameter and $\mathcal{N}_i$ is the negative sample set for item $i$. Essentially, Eq.~(\ref{eq:loss}) is a cross-entropy loss between the positive pair and all other negative pairs. For a predicted vector $\hat{\mathbf{z}}_i$, the only positive pair is $(\hat{\mathbf{z}}_i,\mathbf{z}_i)$ because $\mathbf{z}_i$ is the corresponding ground-truth at time step $i$. Except for the ground truth item, all other latent representations $\mathbf{z}_*$ from any other items consist of the negative pair samples.

\subsubsection{Negative Sampling within Batch}
\label{sec:neg}
Since there is a negative set in the training objective, a sampling procedure of this set is needed. Generally, there are two types of sampling methods for the negative set: memory bank~\cite{moco} and batch sampling~\cite{simclr,cpc,ecpc,memdpc,deepinfomax}. In MMInfoRec, the calculation of the NCE is inside a batch rather than sampling across the whole item set since a batch will contain enough negative samples. In the CPC methods for computer vision tasks~\cite{cpc,ecpc,memdpc,deepinfomax}, the negative sample can be chosen from the channels of the same feature map or other feature maps from the same batch. Similarly, in sequential recommendation, we can use the feature vector $\mathbf{z}$ of other items in the batch to construct the negative sample set $\mathcal{N}_i$ for every item $i$.

\paragraph{General negatives}
For items in other sequences of the same batch, they can generally serve as negative samples of the predicted preference. Since these items have a large potential to represent totally different preferences from other users, they can be considered as easy negatives in $\mathcal{N}_i$ of Eq.~(\ref{eq:milnce}). On the right of Fig.~\ref{fig:whole}, there is a General Neg.\ in the MINCE loss.

\paragraph{Temporal negatives}
For other items except for the ground truth positive sample in the same sequence, they are more difficult for the model to discriminate. Because these items and the ground truth are in the same sequence, these other items have the tendency to represent a outdated and misleading preference of a user. Thus, it is important to include them as hard negatives to train the model, which could enable the model to discriminate the difference. On the right-hand side of Fig.~\ref{fig:whole}, there is a Temporal Neg.\ in the MINCE loss.

\paragraph{Number of negatives}
As described above, the sampling of negative samples is conducted within the training mini batch. Assume there are $D$ unique items in the training batch. According to the negative sampling strategy, the number of samples in $\mathcal{N}_i$ is denoted as $|\mathcal{N}_i|=D-1$.

\subsubsection{Multiple Instance Positive Sampling}
\label{sec:pos}
For the predictive latent representation $\hat{\mathbf{z}}_i$, the most natural choice of positive pair is $\left(\hat{\mathbf{z}}_i,\mathbf{z}_i\right)$, which chooses the ground-truth latent vector $\mathbf{z}_i$ of the corresponding time step. In this section, we will extend this positive sampling strategy to a multiple instance scheme.

To extend the vanilla NCE from single positive sample to multiple instances of semantically positive sample, the key challenge is to identify the semantically positive samples of the ground truth for the next item prediction. Since the sequential recommendation usually considers the next item as the label of prediction, it is generally difficult to find out the semantically similar items as the multiple positive samples in NCE. This difficulty is mainly due to the high unavailability of defining the semantic similarity of sequences, which usually relies on counting on the nearest neighbor in the sequence level~\cite{stan}. Different from data samples in computer vision research field, e.g., images and videos, there is not widely applicable augmentation methods for the embedding representation of items to obtain the semantic positive samples~\cite{moco,simclr} nor meaningful labels to mine the semantic positive samples in a supervised learning style~\cite{scl,milnce}.

To address this difficulty for semantically similar positive samples, a Dropout-based positive sample mining strategy is proposed in MMInfoRec. As described in Section~\ref{sec:attenc}, an item is encoded by $g_\text{enc}$ into a dense embedding vector. When there is any Dropout function inside $g_\text{enc}$, a set of different yet semantically similar encoded vectors of the same item can be obtained by setting different Dropout masks in $g_\text{enc}$ on the same item. Assume there are $q$ different random Dropout functions. When these Dropouts are operated with $g_\text{enc}$ on the same item $i$, a set of different latent representations of the same item are denoted as $\mathcal{P}_i=\{\mathbf{z}^1_i,\mathbf{z}^2_i,\ldots,\mathbf{z}^q_i\}$. These semantic positives are denoted as Sem.\ Pos.\ in Fig.~\ref{fig:whole}. After generating the semantic positive set for every item in the batch, with this $\mathcal{P}_i$ serving as the semantically positive set, the multi-instance NCE (MINCE) loss can include more than one term in the nominator:
\begin{equation}
\label{eq:milnce}
    \ell_\text{MINCE}=-\sum_{i}\left[\log \frac{\sum\limits_{* \in \mathcal{P}_i}e^{\left(\hat{\mathbf{z}}_{i}^{\top} \cdot \mathbf{z}_{*}\right)/\tau}}{\sum\limits_{* \in \mathcal{P}_i}e^{\left(\hat{\mathbf{z}}_{i}^{\top} \cdot \mathbf{z}_{*}\right)/\tau}+\sum\limits_{* \in \mathcal{N}_i}e^{\left(\hat{\mathbf{z}}_{i}^{\top} \cdot \mathbf{z}_{*}\right)/\tau}}\right],
\end{equation}
where $\mathcal{N}_i$ is the negative sample set for item $i$ and naturally contains the the $q$ variants of the items in the negative set in Eq.~(\ref{eq:loss}). Thus, the size of the negative set in Eq.~(\ref{eq:milnce}) is denoted as $|\mathcal{N}_i|=q(D-1)$. In the training of MMInfoRec, $\ell_\text{MINCE}$ is chosen to be the training objective.

\subsection{Discussion}
\label{sec:discuss}
Our MMInfoRec model provides a novel contrastive training scheme in sequential recommendation task. Fot the previous methods, the most common training scheme is the next-item supervised training. Recent state-of-the-art methods, for example, GRU4Rec~\cite{gru4rec}, SASRec~\cite{sasrec}, Caser~\cite{caser}, FDSA~\cite{fdsa} and Seq2SeqRec~\cite{s2s}, all applies this scheme to train their models. contrastive learning is introduced to this field recently by adding a pretext task to train an encoder to learn a general understanding of the interaction sequence. BERT4Rec~\cite{bert4rec} and $\text{S}^3$Rec both utilize the masked training from masked language models. They rely on representing the item with its neighbors. Such a training style in natural language understanding can improve the learned word embedding to generalize to different downstream tasks. However, in the sequential recommendation, there is only one desired task, the recommendation. Therefore, both of these methods need a further finetuning step to make the model target at the recommendation. In contrast, the proposed MMInfoRec have a consensus in the pretext task and the recommendation task by ranking the positive sample in a higher position than negative samples, which prevents from a finetuning after the contrastive pre-training.

This vanilla NCE, $\ell_\text{NCE}$ in Eq.~(\ref{eq:loss}) indicates the model to assign a higher similarity between representations from positive pairs than other negative samples. Under this situation, the model can distinguish the positive predictive result from the whole item set. In general, $\ell_\text{NCE}$ is similar to the BPR loss~\cite{bprmf} while using multiple negative samples. Compared with listwise losses, $\ell_\text{NCE}$ focuses more on ranking the target item in a higher position than all the negative samples while listwise losses will pay attention to the ranking of negative samples. The MINCE loss $\ell_\text{MINCE}$ proposed Eq.~(\ref{eq:milnce}) offers a novel perspective of the contrastive learning of sequential recommendation. The general NCE only considers the item itself as a positive sample. However, as pointed out in the literature of supervised learning with contrastive signals, when semantically similar samples are included into the contrastive training, the encoding result can be more accurate~\cite{coclr,scl,milnce}. Our proposed MINCE shares the same spirit as these methods to avoid inappropriate categorization of positive and negative samples.

\section{Experiment}
\label{sec:exp}
In this section, we will describe the experiments to verify the efficacy of the MMInfoRec. In Section~\ref{sec:setup}, the experimental setup is demonstrated. In the following sections, we will answer research questions (RQ) by different experiments:

\begin{itemize}
    \item \textbf{RQ1}: How does the MMInfoRec perform compared with the state-of-the-art sequential recommender models? (Section~\ref{sec:overall-exp})
    \item \textbf{RQ2}: How does each proposed component contribute to the whole model? (Section~\ref{sec:ablation})
    \item \textbf{RQ3}: How does the memory module improve the sequential recommendation? (Section~\ref{sec:mem-exp})
    \item \textbf{RQ4}: How does MINCE contribute to the contrastive training in the sequential recommendation? (Section~\ref{sec:nce-exp})
    \item \textbf{RQ5}: How is the sensitivity of the hyper-parameters of the MMInfoRec model? (Section~\ref{sec:param-exp})
\end{itemize}

\subsection{Setup}
\label{sec:setup}
In this section, the setup of experiments is presented, including datasets (Section~\ref{sec:datasets}) with the corresponding preprocessing (Section~\ref{sec:prepro}), metrics used to evaluate the performance (Section~\ref{sec:metric}), state-of-the-art baselines (Section~\ref{sec:baseline}) and the implementation (Section~\ref{sec:imple}).

\begin{table}[t]
    \caption{Statistics of the datasets after preprocessing.}
    \begin{tabular}{l|rrrr}
    \toprule
    & Beauty & Sports & Toys & Yelp \\
    \midrule
    $\sharp$ Users & 22,363 & 35,598 & 19,412 & 30,431\\
    $\sharp$ Items & 12,101 & 18,357 & 11,924 & 20,033\\
    $\sharp$ Avg. Actions / User & 8.9 & 8.3 & 8.6 & 10.4\\
    $\sharp$ Avg. Actions / Item & 16.4 & 16.1 & 14.1 & 15.8\\
    $\sharp$ Actions & 198,502 & 296,337 & 167,597 & 316,354\\
    Sparsity & 99.93\% & 99.95\% & 99.93\% & 99.95\%\\
    $\sharp$ Attributes & 1,221 & 2,277 & 1,027 & 1,001\\
    $\sharp$ Avg. Attribute / Item & 5.1 & 6.0 & 4.3 & 4.8 \\
    \bottomrule
    \end{tabular}
    \label{tab:datasets}
\end{table}

\subsubsection{Datasets}
\label{sec:datasets}
To evaluate the proposed MMInfoRec model, the following datasets are applied with a summary in Table~\ref{tab:datasets}:
\begin{itemize}
    \item \textbf{Amazon Beauty, Sports, and Toys}~\cite{amazon}\footnote{http://jmcauley.ucsd.edu/data/amazon/}. Following~\cite{bert4rec,s3rec,s2s}, we choose three subcategories of the Amazon dataset with their fine-grained categories and the brands of the product as attributes.
    \item \textbf{Yelp}\footnote{https://www.yelp.com/dataset}. Yelp is a widely used dataset for the business recommendation. Similar to~\cite{s3rec}, the transaction records after Jan. 1st, 2019 are used in our experiment. The categories of businesses are considered as attributes.
\end{itemize}

\subsubsection{Preprocessing}
\label{sec:prepro}
For all datasets, we treat all the interaction records from the same user as a sequence. The order is given by the timestamp of the interaction. Following~\cite{s3rec,bert4rec,fdsa}, item appearing frequencies less than 5 will be filtered out. And if a sequence is shorter than 5, the sequence will also be removed. The maximum length is set to 50, which means that only the most recent 50 interactions will be kept if the sequence is longer than 50 interactions. The second last item in a sequence is used for validation and the last item is used for testing. The rest items are used for training.

\subsubsection{Metrics}
\label{sec:metric}
To evaluate the performance by fair comparisons, top-$K$ Hit Ratio (HR@$K$) and top-$K$ Normalized Discounted Cumulative Gain (NDCG@$K$) are applied with $K$ chosen from $\{5,10\}$ following previous methods~\cite{bert4rec,s3rec,s2s}. These usually apply a sampling strategy for evaluation, which is proved to be unfair by recent work~\cite{metric}. Therefore, we evaluate the ranking result over the whole item set.

\begin{table*}[t]
    \caption{Overall performance. Bold scores represent the highest results of all methods. Underlined scores stand for the highest results from previous methods. The MMInfoRec achieves the state-of-the-art result among all baseline models.}
    \resizebox{\linewidth}{!}{
    \begin{tabular}{c|l|ccccccccc|c|c}
    \toprule
         Dataset& Metric & GRU4Rec & Caser & SASRec & BERT4Rec & HGN&$\text{GRU4Rec}_F$&$\text{SASRec}_F$&FDSA&$\text{S}^3\text{Rec}$ &MMInfoRec& Improv.\\
         \midrule
         \multirow{4}*{Beauty}&HR@5&0.0164 & 0.0205 & \underline{0.0387} & 0.0203 & 0.0325 & 0.0248&0.0268 & 0.0267 & \underline{0.0387}&\textbf{0.0523}$\pm$0.0007&35.14\%\\
         &HR@10&0.0283 & 0.0347 & 0.0605 & 0.0347 & 0.0512 & 0.0420&0.0458 & 0.0407 & \underline{0.0647}&\textbf{0.0730}$\pm$0.0010&12.83\%\\
         &NDCG@5&0.0099 & 0.0131 & \underline{0.0249} & 0.0124 & 0.0206 & 0.0156&0.0171 & 0.0163 & 0.0244&\textbf{0.0378}$\pm$0.0004&51.81\%\\
         &NDCG@10&0.0137 & 0.0176 & 0.0318 & 0.0170 & 0.0266 & 0.0211&0.0232 & 0.0208 &\underline{0.0327}&\textbf{0.0445}$\pm$0.0003&36.09\%\\
         \midrule
         \multirow{4}*{Sports}&HR@5&0.0129 & 0.0116 & 0.0233 & 0.0115 & 0.0189 & 0.0153&0.0163 & 0.0182 &\underline{0.0251}&\textbf{0.0274}$\pm$0.0009&9.16\%\\
         &HR@10&0.0204 & 0.0194 & 0.0350 & 0.0191 & 0.0313 & 0.0260 & 0.0277 & 0.0288 & \underline{0.0385}&\textbf{0.0398}$\pm$0.0009&3.38\%\\
         &NDCG@5&0.0086 & 0.0072 & 0.0154 & 0.0075 & 0.0120 & 0.0102 & 0.0107 & 0.0122 & \underline{0.0161}&\textbf{0.0196}$\pm$0.0005&21.74\%\\
         &NDCG@10&0.0110 & 0.0097 & 0.0192 & 0.0099 & 0.0159 & 0.0136 & 0.0144 & 0.0156 & \underline{0.0204}&\textbf{0.0231}$\pm$0.0006&13.24\%\\
         \midrule
         \multirow{4}*{Toys}&HR@5&0.0097 & 0.0166 & \underline{0.0463} & 0.0116 & 0.0321 & 0.0268 & 0.0289 & 0.0228 & 0.0443&\textbf{0.0611}$\pm$0.0005&31.97\%\\
         &HR@10&0.0176 & 0.0270 & 0.0675 & 0.0203 & 0.0497 & 0.0427 & 0.0439 & 0.0381 & \underline{0.0700}&\textbf{0.0813}$\pm$0.0009&16.14\%\\
         &NDCG@5&0.0059 & 0.0107 & \underline{0.0306} & 0.0071 & 0.0221 & 0.0179 & 0.0194 & 0.0140 & 0.0294&\textbf{0.0451}$\pm$0.0004&47.34\%\\
         &NDCG@10&0.0084 & 0.0141 & 0.0374 & 0.0099 & 0.0277 & 0.0230 & 0.0242 & 0.0189 & \underline{0.0376}&\textbf{0.0516}$\pm$0.0004&37.23\%\\
         \midrule
         \multirow{4}*{Yelp}&HR@5&0.0152 & 0.0151 & 0.0162 & 0.0051 & 0.0186 & 0.0176 & 0.0170 & 0.0158 & \underline{0.0201}&\textbf{0.0508}$\pm$0.0005&152.74\%\\
         &HR@10&0.0263 & 0.0253 & 0.0274 & 0.0090 & 0.0326 & 0.0285 & 0.0284 & 0.0276 & \underline{0.0341} & \textbf{0.0615}$\pm$0.0008&80.35\%\\
         &NDCG@5&0.0099 & 0.0096 & 0.0100 & 0.0033 & 0.0115 & 0.0110 & 0.0110 & 0.0098 & \underline{0.0123}&\textbf{0.0327}$\pm$0.0012&165.85\%\\
         &NDCG@10&0.0134 & 0.0129 & 0.0136 & 0.0045 & 0.0159 & 0.0145 & 0.0147 & 0.0136 & \underline{0.0168}&\textbf{0.0367}$\pm$0.0004&118.45\%\\
         \bottomrule
    \end{tabular}
    }
    \label{tab:overall}
\end{table*}

\subsubsection{Baselines}
\label{sec:baseline}
We choose the baselines according to current research~\cite{s3rec,fdsa} and utilize the most popular and state-of-the-art methods for comparisons. We omit shallow methods and non-sequential methods since they are proved to be unable to show competitive results with recent neural-based sequential recommender models~\cite{s3rec,fdsa}. The following baselines are chosen to compare with the proposed MMInfoRec:
\begin{itemize}
    \item \textbf{GRU4Rec}~\cite{gru4rec} applies GRU to model user sequences. A sequence is treated as the input and the output hidden state serves as the sequence representation.
    \item \textbf{Caser}~\cite{caser} is a CNN-based method capturing high-order Markov Chains by applying horizontal and vertical convolutional operations for the sequential recommendation. This structure is majorly different from usual sequence modeling methods in the choice of network type.
    \item \textbf{SASRec}~\cite{sasrec} is a single-directional self-attention based sequential recommendation model, which uses the multi-head attention mechanism to recommend the next item. It is a strong baseline in sequential recommendation since the attention mechanism is a powerful tool in sequence modeling, which is also demonstrated in NLP.
    \item \textbf{BERT4Rec}~\cite{bert4rec} uses a masked item training scheme similar to the masked language model. The backbone is the bi-directional self-attention mechanism.
    \item \textbf{HGN}~\cite{hgn} is recently proposed and adopts hierarchical gating networks to capture long-term and short-term user interests. User information is used for both the gating network and the final recommendation score.
    \item \textbf{$\text{GRU4Rec}_F$}~\cite{gru4recf} is an improved version of GRU4Rec, which leverages attributes. This model calculates the sequence representation in a parallel style.
    \item \textbf{$\text{SASRec}_F$} is an extension of SASRec from previous work~\cite{s3rec}, which concatenates the representations of items and attribute as the input to the model.
    \item \textbf{FDSA}~\cite{fdsa} constructs a feature sequence and uses a feature level self-attention block to model the feature transition patterns. It is a two-stream model that calculates the representation of a sequence based on the ID information and side information of items respectively.
    \item \textbf{S$^3$Rec}~\cite{s3rec} applied masked contrastive pre-training similar to BERT4Rec. The mask is utilized on segments, attributes and single item. It serves as a strong baseline for sequential recommendation using side information of items. Recent Seq2SeqRec~\cite{s2s} method uses a similar next sequence prediction, which is a special case of S$^3$Rec.
\end{itemize}
The full ranking results of these baselines are from the updated result\footnote{https://github.com/aHuiWang/CIKM2020-S3Rec} provided by $\text{S}^3\text{Rec}$, which will be shown in Table~\ref{tab:overall}.

\subsubsection{Implementation}
\label{sec:imple}
In Section~\ref{sec:method}, we provide the overall framework of the MMInfoRec. For the detailed choice of each module, we describe each module in this section. For the attribute encoder $g_\text{enc}$, we choose to use the Transformer~\cite{attention} module to perform a self-attention of the item embedding to all of its attribute embeddings. For the temporal aggregation $g_\text{ar}$, another Transformer~\cite{attention} is applied. Meanwhile, a learned positional encoding layer is added to the input of $g_\text{ar}$. For the auto-regressive predictive function $g_\text{ap}$, a GRU network~\cite{gru} is used for the multi-step predictive computation.

In this paragraph, we will describe the hyper-parameter setting for the training and the network. The embedding size is set to $64$ with all linear mapping function in the MMInfoRec has the same hidden size. The number of layer and head in the Transformer are chosen from $\{1,2\}$. A Dropout~\cite{dropout} function with the ratio $0.5$ is used on both the input of the $g_\text{ta}$ function and the Transformer module inside $g_\text{ta}$ to alleviate the overfitting issue. The training batch size is set to $256$. We use the Adam~\cite{adam} optimizer with the learning rate chosen from $\{0.0003,0.001,0.003,0.01,0.03\}$. The number of memory slot $b$ is chosen from $\{5,10,32,64,128,256\}$. The default number of the predictive step is chosen from $\{1,2,3,4\}$. The number of different Dropout functions in Eq.~(\ref{eq:milnce}) is chosen from $\{1,2,3,4\}$. The temperature parameter $\tau$ in Eq.~(\ref{eq:milnce}) is chosen from $\{0.1,0.3,0.6,1,3\}$. An $\ell_2$ regularisation is also applied along with the training, with the weight chosen from $\{0,0.1,0.01,0.001,0.0001,0.00001\}$.

\subsection{Overall Performance}
\label{sec:overall-exp}
In this section, we compare MMInfoRec with state-of-the-art methods. We evaluate the performance on four datasets, Amazon Beauty, Amazon Sports, Amazon Toys and Yelp. The metrics used in this experiment are HR@5, NDCG@5, HR@10 and NDCG@10. The experimental result is shown in Table~\ref{tab:overall}.

According to the result in Table~\ref{tab:overall}, the MMInfoRec achieved the best performance compared with all the baseline methods. The improvement percentage is significant across all datasets.

For sequential recommendation baseline methods without using the side information, they can achieve comparable results in certain situations. However, most of the time, they are far away from the highest performance. For example, GRU4Rec is the earliest neural-based method in the sequential recommendation. GRU4Rec utilizes GRU to encode the sequence. It generally has a less competitive result. Similar performance is achieved by Caser, which uses a convolutional neural network to compute the representation of a sequence. The backbone of these methods is not strong enough to help boost the experimental performance of models. SASRec and BERT4Rec use a single-directional and a bi-directional attention mechanism respectively to learn the dependence between items in a sequence. SASRec can serve as a strong baseline in this task since it can sometimes beat the methods that rely on extra item features. However, BERT4Rec is not as strong as SASRec, which could be because the indirect training scheme of BERT4Rec introduces unrelated information into the representation learning of items. However, they still prove that the Transformer structure is a strong backbone in the sequential recommendation. HGN includes the user information to support a structural gating network to learn the interaction between the user information and the item information. It can achieve comparable results with SASRec although not using a Transformer structure. This situation demonstrates that a carefully designed hierarchical gating network can model the relationship between items well.

For the methods that using item attributes in the sequential recommendation, the performance of them is different. $\text{GRU4Rec}_F$ and $\text{SASRec}_F$ are two extensions to the original non-feature-based methods GRU4Rec and SASRec respectively. They achieve similar results that outperform the basic GRU4Rec method. However, their performance is still interior to SASRec, which indicates that an inappropriate inclusion of the side information would hurt the performance of the model itself. A simple concatenation of the item's ID representation and its attributes representations are unable to fuse the information effectively. FDSA is a method that mixes the parallel structure of $\text{GRU4Rec}_F$ with the Transformer network of SASRec to encode the ID embedding and attribute embeddings separately. The performance of FDSA is then similar to the feature-based methods since it uses the same next-item supervised training scheme with a similar network structure. The strongest baseline of the sequential recommendation is S$^3$Rec, which uses a masked sequence modeling simultaneously on the ID embedding, attribute embeddings and segment representations. The contrastive training scheme benefits the model a lot by injecting the side information and the prospective intention into ID embeddings. It achieves the best result in most of the situations compared with other baselines. It proves that contrastive learning could be a promising technique in sequential recommendation tasks.

Given the overall performance, we can inspect the reasons for the improvement of the MMInfoRec over these baselines. (1) Compared with the traditional sequential recommender systems, GRU4Rec, Caser, SASRec and HGN, the MMInfoRec exploits the attributes of items to assist the learning of representations of items and sequences. (2) Compared with the feature-based sequential recommender systems, $\text{GRU4Rec}_F$, $\text{SASRec}_F$, FDSA and S$^3$Rec, the MMInfoRec uses a self-attention structure to fuse the attributes of an item, which enables the model to learn the interaction between different attributes and the ID information. (3) Compared with the contrastive methods, BERT4Rec and S$^3$Rec, the MMInfoRec focuses on a pretext task that is more similar to the recommendation task itself. In contrast, the pretext tasks in BERT4Rec and S$^3$Rec are masked sequential prediction, which has a gap to our desired recommendation task. Therefore, the MMInfoRec has a more effective way to make use of the training signal.

\begin{table}[t]
    \caption{Ablation study for the memory module and the MINCE objective.}
    \resizebox{\linewidth}{!}{
    \begin{tabular}{c|l|ccc|c}
    \toprule
         Dataset& Metric &CPC& +$g_\text{m}$ & +MINCE &MMInfoRec\\
         \midrule
         \multirow{4}*{Beauty}&HR@5&0.0432$\pm$0.0009&0.0469$\pm$0.0008 & 0.0515$\pm$0.0007 &0.0523$\pm$0.0007\\
         &NDCG@5&0.0310$\pm$0.0004&0.0346$\pm$0.0005 & 0.0364$\pm$0.0004 &0.0378$\pm$0.0004\\
         &HR@10&0.0643$\pm$0.0013&0.0665$\pm$0.0009 & 0.0708$\pm$0.0009 &0.0730$\pm$0.0010\\
         &NDCG@10&0.0362$\pm$0.0014&0.0408$\pm$0.0012 & 0.0424$\pm$0.0006 &0.0445$\pm$0.0003\\
         \midrule
         \multirow{4}*{Sports}&HR@5&0.0223$\pm$0.0005 &0.0245$\pm$0.0008& 0.0251$\pm$0.0013&0.0274$\pm$0.0009\\
         &NDCG@5&0.0161$\pm$0.0009&0.0177$\pm$0.0006 & 0.0183$\pm$0.0007&0.0196$\pm$0.0005\\
         &HR@10&0.0317$\pm$0.0012&0.0332$\pm$0.0007 & 0.0372$\pm$0.0015&0.0398$\pm$0.0009\\
         &NDCG@10&0.0182$\pm$0.0008&0.0206$\pm$0.0012 & 0.0215$\pm$0.0006&0.0231$\pm$0.0006\\
         \midrule
         \multirow{4}*{Toys}&HR@5&0.0549$\pm$0.0015& 0.0586$\pm$0.0009 & 0.0587$\pm$0.0014&0.0611$\pm$0.0005\\
         &NDCG@5&0.0415$\pm$0.0005&0.0436$\pm$0.0012 & 0.0435$\pm$0.0009 &0.0451$\pm$0.0004\\
         &HR@10&0.0737$\pm$0.0006&0.0792$\pm$0.0008 & 0.0795$\pm$0.0017&0.0813$\pm$0.0009\\
         &NDCG@10&0.0469$\pm$0.0004&0.0496$\pm$0.0011 & 0.0494$\pm$0.0013&0.0516$\pm$0.0004\\
         \midrule
         \multirow{4}*{Yelp}&HR@5&0.0396$\pm$0.0007&0.0462$\pm$0.0009 & 0.0481$\pm$0.0010&0.0508$\pm$0.0005\\
         &NDCG@5&0.0221$\pm$0.0005 & 0.0283$\pm$0.0008 &0.0304$\pm$0.0005&0.0327$\pm$0.0012\\
         &HR@10&0.0526$\pm$0.0008&0.0581$\pm$0.0006 & 0.0597$\pm$0.0006 &0.0615$\pm$0.0008\\
         &NDCG@10&0.0265$\pm$0.0006 & 0.0331$\pm$0.0012&0.0352$\pm$0.0007&0.0367$\pm$0.0004\\
         \bottomrule
    \end{tabular}
    }
    \label{tab:ablation}
\end{table}

\subsection{Ablation Study}
\label{sec:ablation}
In this experiment, we conduct the ablation study to verify the efficacy of each components in MMInfoRec, mainly for the memory module $g_\text{m}$ and the MINCE objective. When both of these two major components are removed from MMInfoRec, the model is degraded to the standard CPC scheme. To verify each of these components, we adding $g_\text{m}$ and MINCE separately to the basic CPC model, denoted as +$g_\text{m}$ and +MINCE respectively. The experimental result is shown in Table~\ref{tab:ablation}.

According to the result, it can be seen that both $g_\text{m}$ and MINCE can consistently improve the recommendation performance. For the performance on the Beauty and the Sports datasets, the contribution of MINCE is more significant than $g_\text{m}$. While for the Toys and Yelp datasets, the contributions of both components are close in terms of the performance. When combining them together into MMInfoRec, the highest performance can always be achieved across all datasets.

\begin{figure}[t]
    \centering
    \subfigure[HR@K on Beauty.]{
    \label{fig:mem-beauty-hr}
    \includegraphics[width=0.465\linewidth]{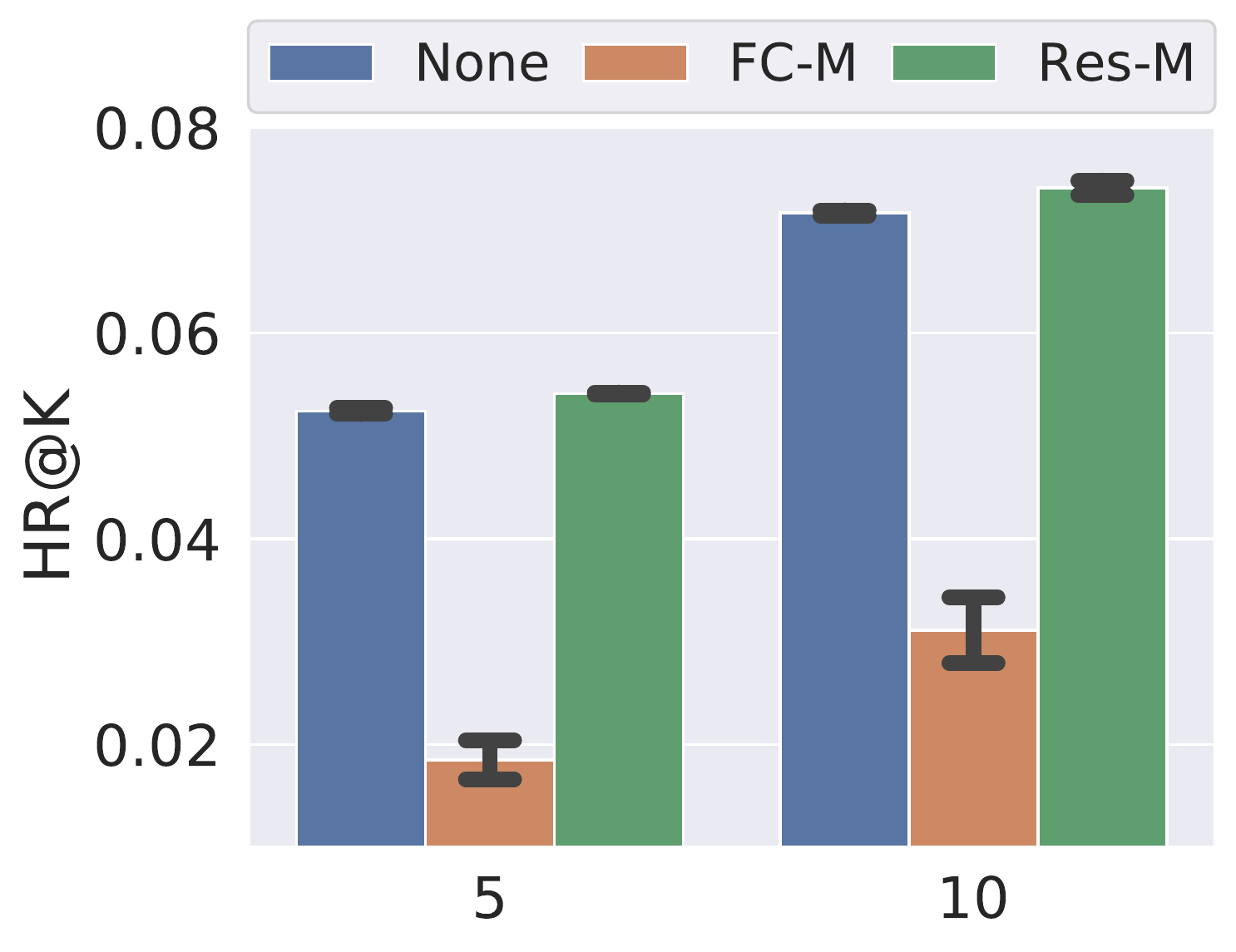}
    }
    \subfigure[NDCG@K on Beauty.]{
    \label{fig:mem-beauty-ndcg}
    \includegraphics[width=0.465\linewidth]{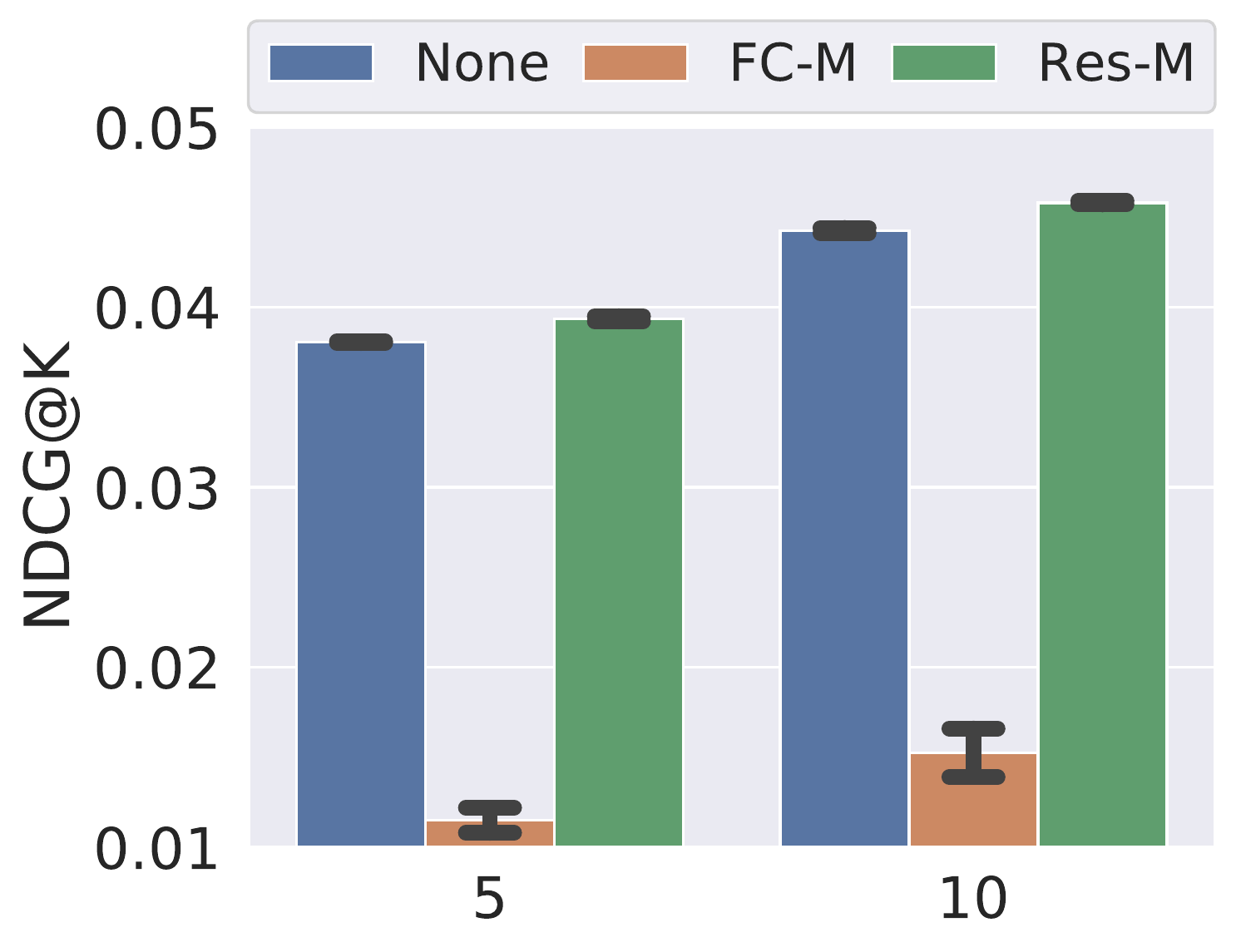}
    }
    \subfigure[HR@K on Toys.]{
    \label{fig:mem-toys-hr}
    \includegraphics[width=0.465\linewidth]{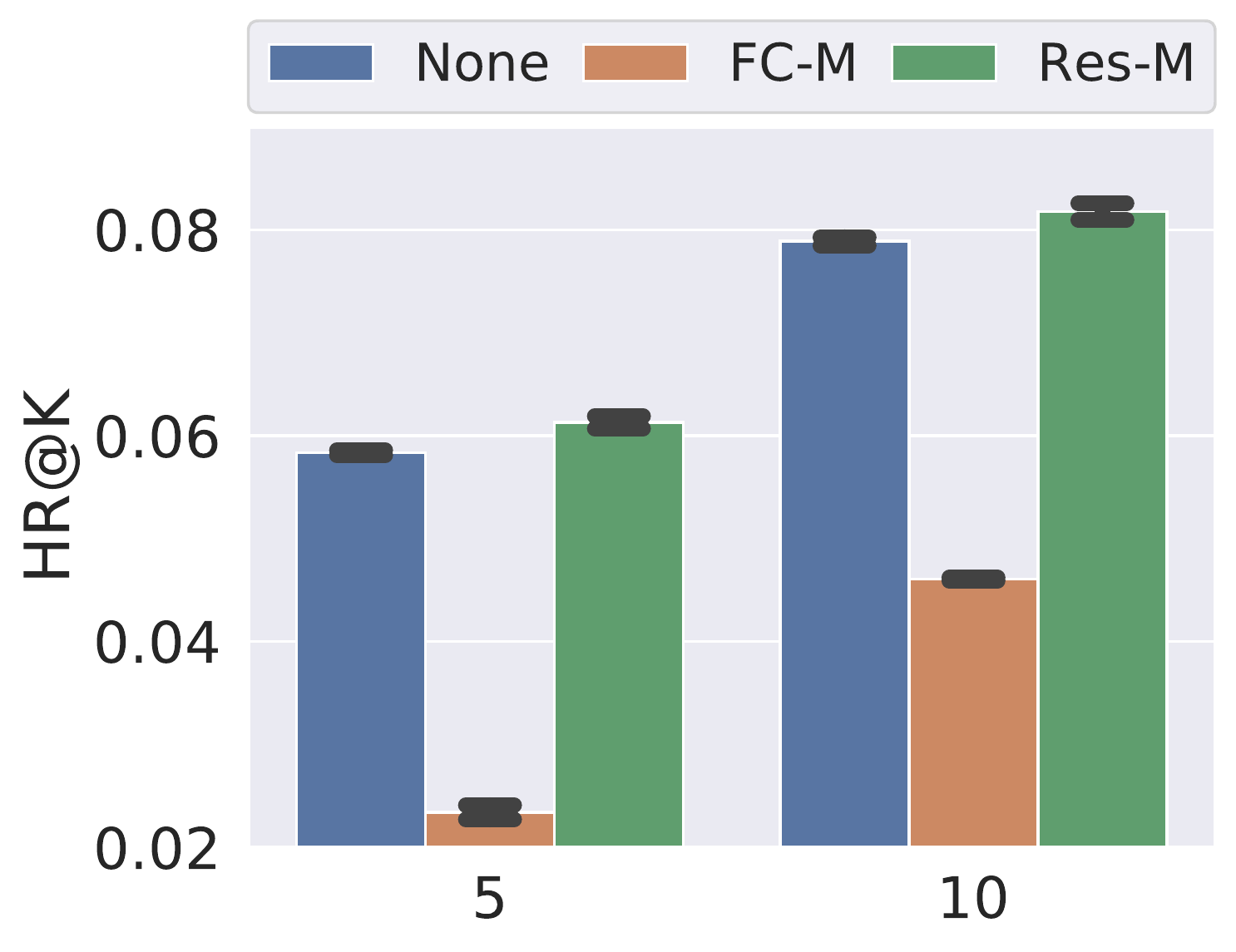}
    }
    \subfigure[NDCG@K on Toys.]{
    \label{fig:mem-toys-ndcg}
    \includegraphics[width=0.465\linewidth]{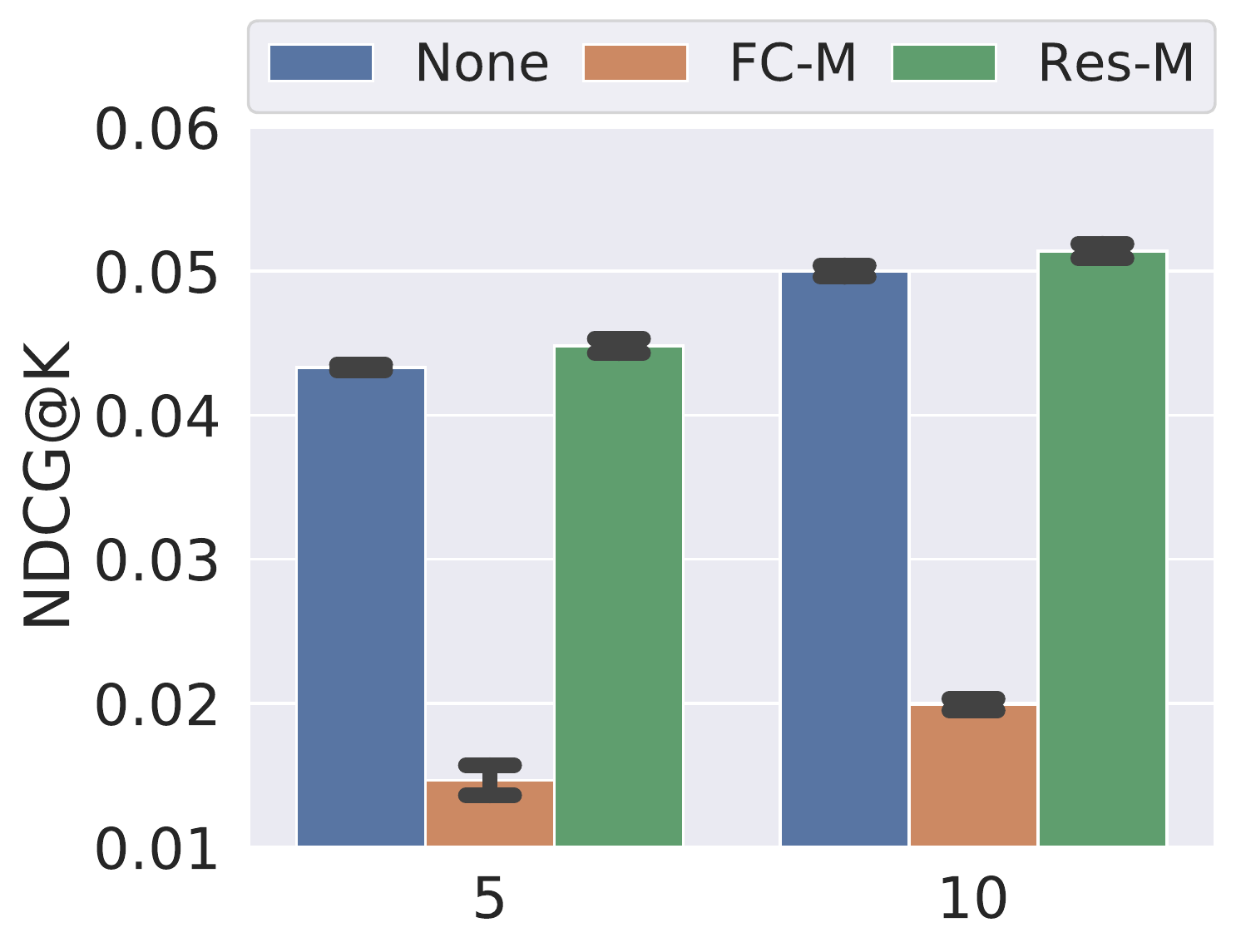}
    }
    \caption{Results of different designs of $g_\text{m}$.}
    \label{fig:mem-design}
\end{figure}

\subsection{Memory Module}
\label{sec:mem-exp}
In this experiment, we aim to investigate how the memory module improves the sequential recommendation. Firstly, the following variants of $g_\text{m}$ will be used: (1) None: the baseline without any memory module; (2) FC-M: by removing the addition of the original context vector $\mathbf{c}_t$ in Eq.~(\ref{eq:mem}); and (3) Res-M: the default memory module with the residual addition as Eq.~(\ref{eq:mem}) denoted. Besides, a visualization of the learned memory will be demonstrated.

\subsubsection{Different designs of $g_\text{m}$}
We compare the performance of the variants above by conducting on Beauty and Toys datasets. The experimental results are shown in Fig.~\ref{fig:mem-design}. It can be seen that Res-M can achieve the highest performance across all situations while the FC-M variant cannot have a comparative result with both the None variant and the Res-M variant. A proper way to use the module is required in the recommendation scenario, which maybe due to the embedding space of items is not similar to the feature-rich content space in computer vision. Solely relying on the memory module to accurately represent the feature space is too difficult for a recommendation model. Therefore, a residual addition is a better choice for the memory.

\begin{figure}[t]
    \centering
    \subfigure[Beauty with 64 slots.]{
    \label{fig:mem-beauty-vis}
    \includegraphics[width=0.465\linewidth]{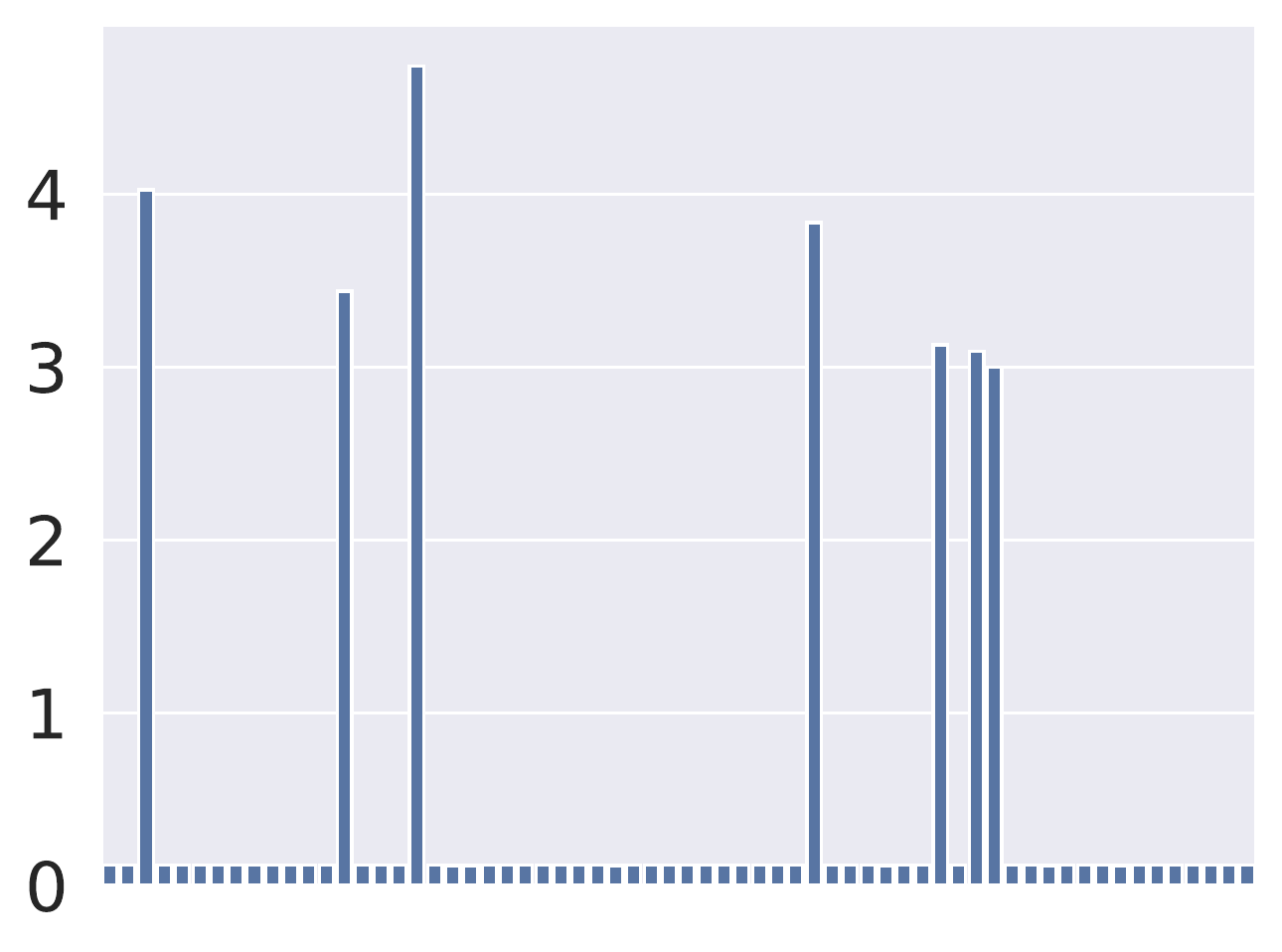}
    }
    \subfigure[Sports with 10 slots.]{
    \label{fig:mem-sports-vis}
    \includegraphics[width=0.465\linewidth]{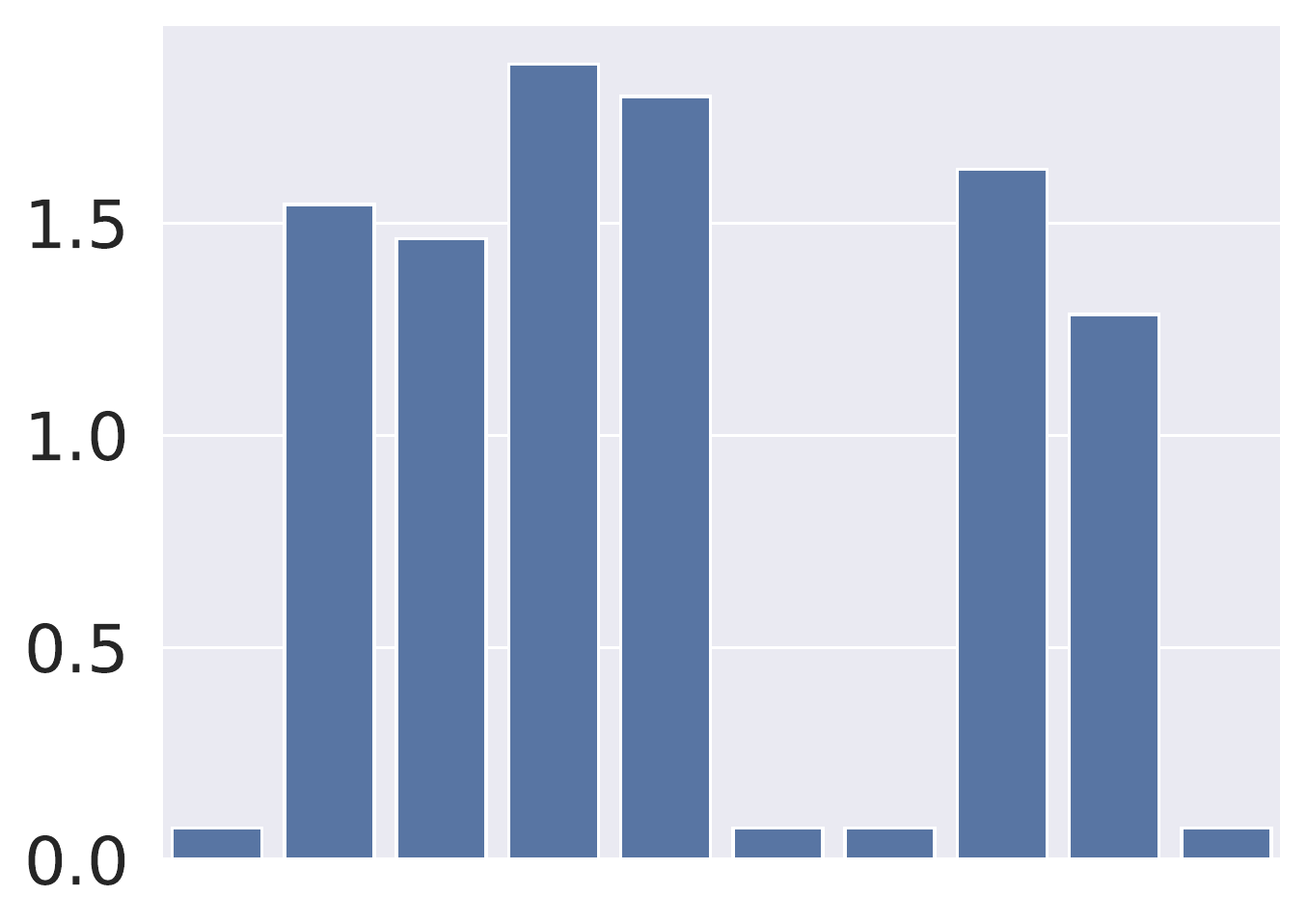}
    }
    \caption{Visualisation of the norm of the memory learned by MMInfoRec on Beauty with 64 slots and Sports with 10 slots.}
    \label{fig:mem-vis}
\end{figure}

\subsubsection{Visualisation of memory}
To illustrate the memory module, the norm of the learned memory is used to demonstrate what is learned within the memory module. The visulisation result is demonstrated in Fig.~\ref{fig:mem-vis}.

It is worth noting that the learned memory bank is a sparse result. For example, in the experiment on Beauty with 64 memory slots, there only seven slots have significant contribution while the rest has a nearly zero norm. Similar situation is shown on the Sports with 10 memory slots, which has six slots contributing profoundly to the model.

\begin{figure}[t]
    \centering
    \subfigure[HR@K on Sports.]{
    \includegraphics[width=0.465\linewidth]{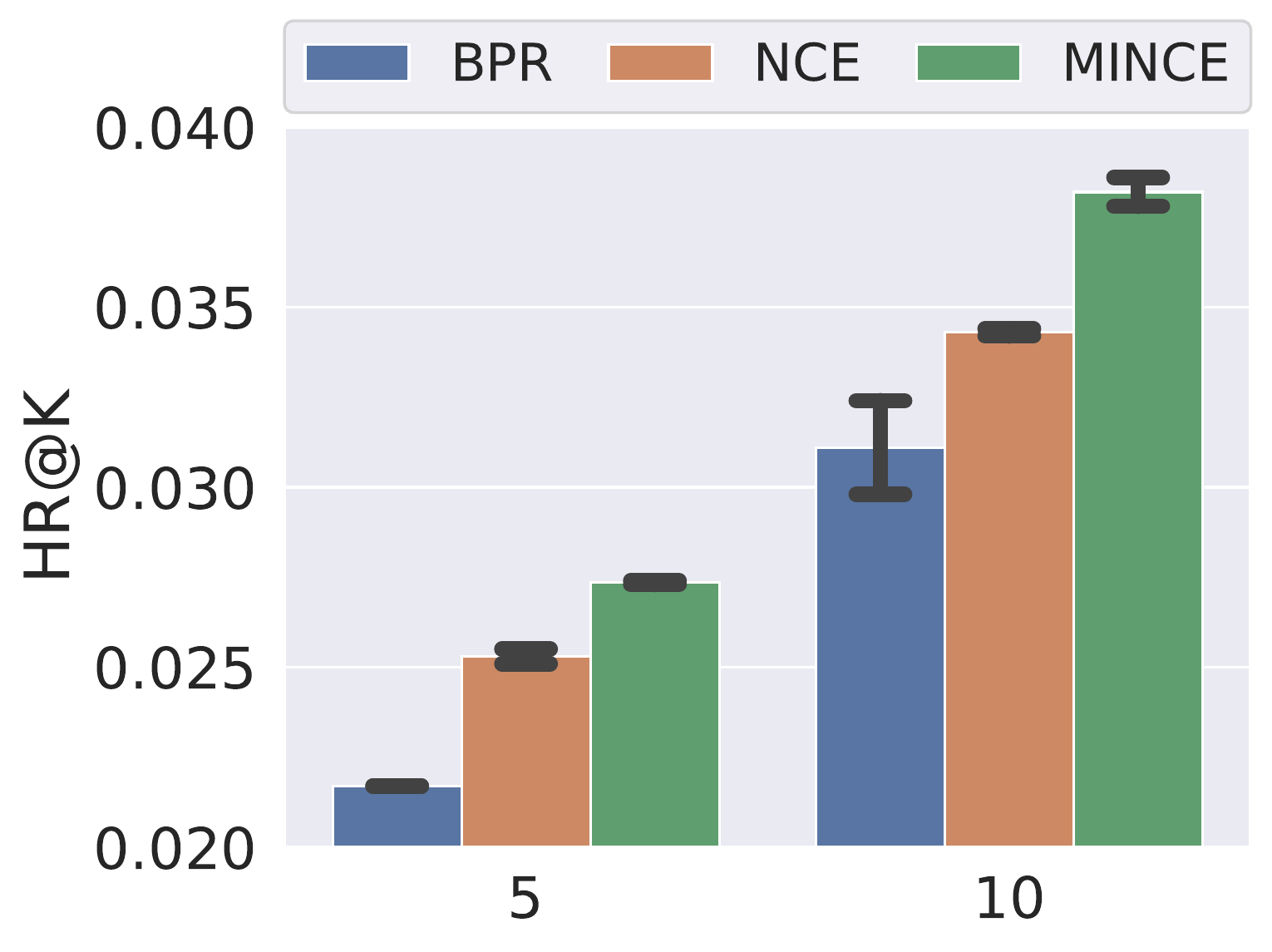}
    }
    \subfigure[NDCG@K on Sports.]{
    \includegraphics[width=0.465\linewidth]{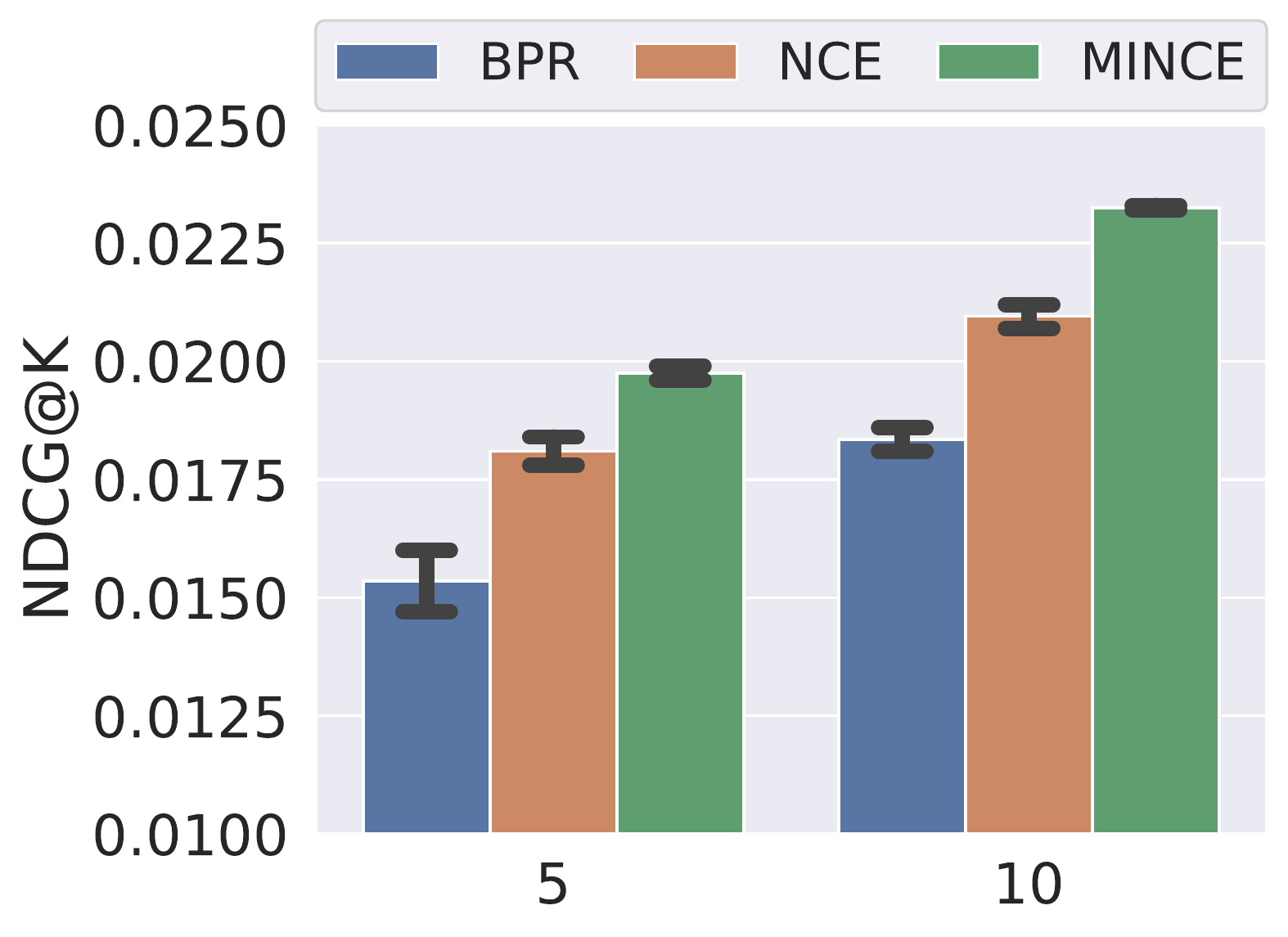}
    }
    \subfigure[HR@K on Toys.]{
    \includegraphics[width=0.465\linewidth]{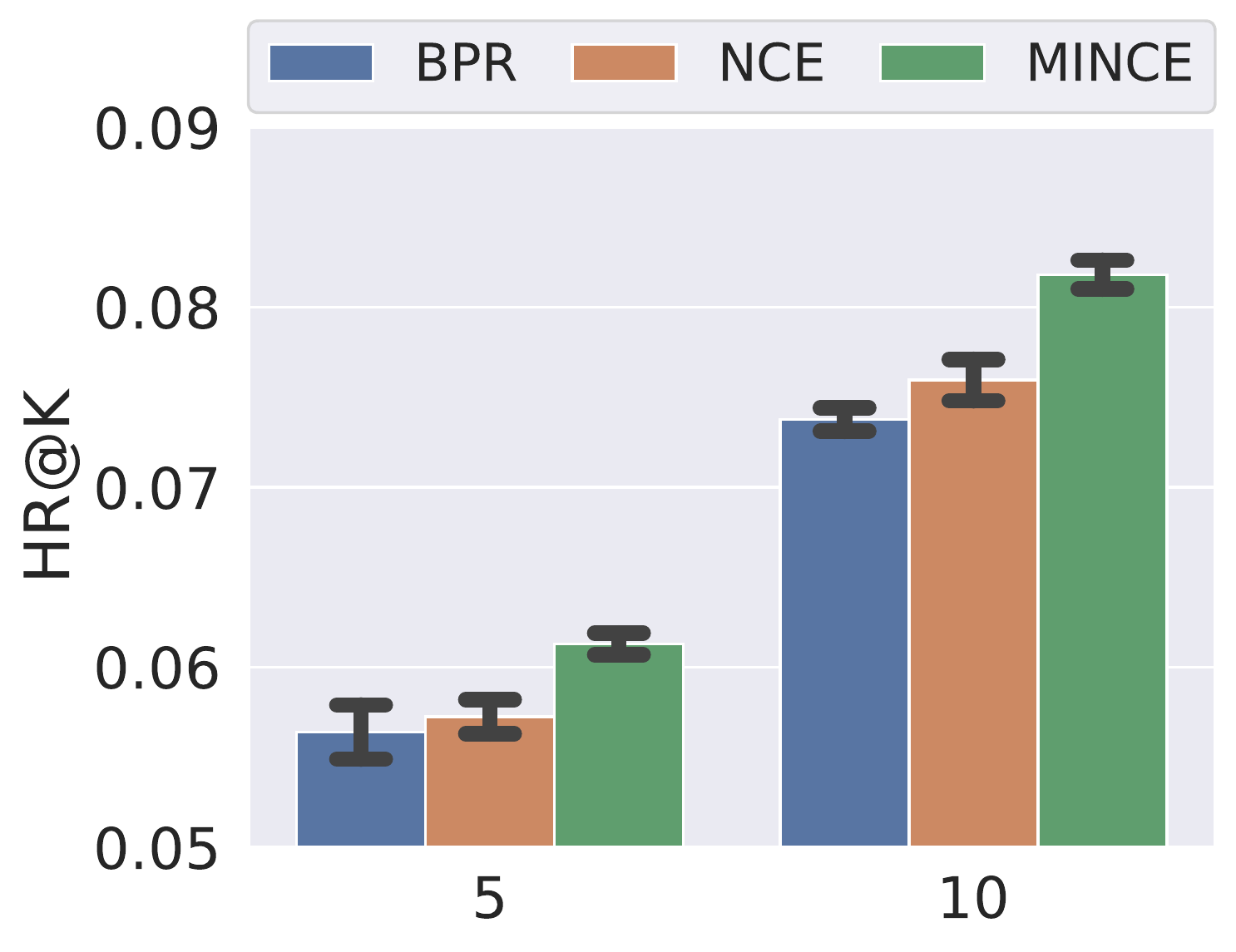}
    }
    \subfigure[NDCG@K on Toys.]{
    \includegraphics[width=0.465\linewidth]{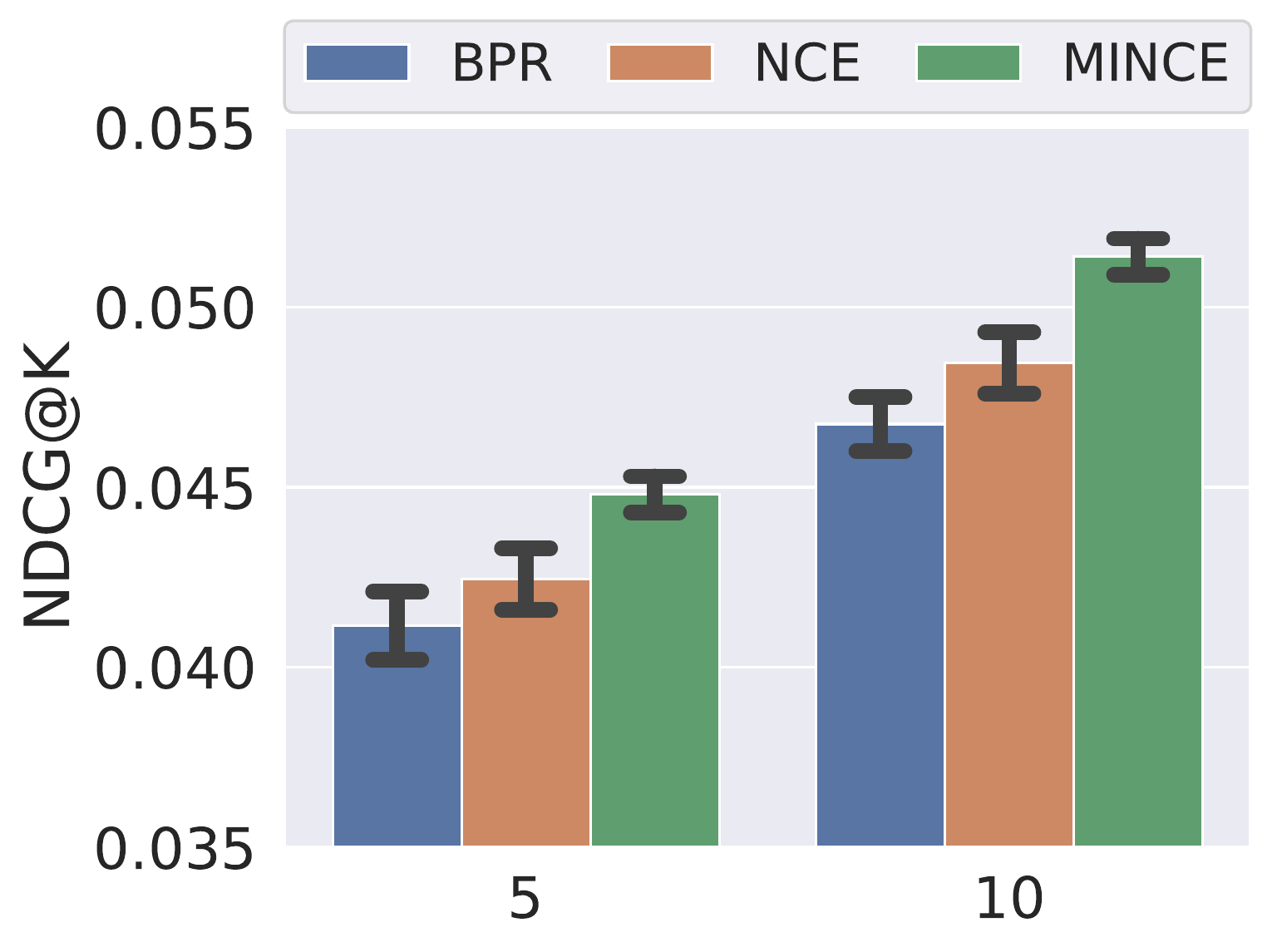}
    }
    \caption{Performance of different ranking objectives.}
\label{fig:nce}
\end{figure}

\subsection{MINCE in Sequential Recommendation}
\label{sec:nce-exp}
In this section, we conduct experiments to evaluate the MINCE loss in Eq.~(\ref{eq:milnce}) in the sequential recommendation. We mainly investigate the vanilla NCE loss in Eq.~(\ref{eq:loss}) and Bayesian pairwise ranking (BPR) loss. The result is shown in Fig.~\ref{fig:nce}.

From Fig.~\ref{fig:nce}, MINCE can always achieve the highest score compared with NCE and BPR. NCE can consistently outperform BPR across all situations. The BPR loss computes the relative preference between the target item and a sampled item. Compared with NCE, BPR only considers one pair of positive and negative items. The estimation of the mutual information between the predicted result and the ground truth for NCE is theoretically proved to be a tight bound. MINCE can be seen as a multi-instance version for the estimation of the mutual information, where the samples are effectively used.

\subsection{Parameter Sensitivity}
\label{sec:param-exp}
There are important hyper-parameters in the MMInfoRec model, for example, the number of memory slot, $b$ in Eq.~(\ref{eq:mem}), the temperature parameter in Eq.~(\ref{eq:milnce}) and the number of the prediction step. In this section, we will evaluate the parameter sensitivity on these hyper-parameters of MMInfoRec.

\begin{figure}[t]
    \centering
    \subfigure[HR@5.]{
    \includegraphics[width=0.465\linewidth]{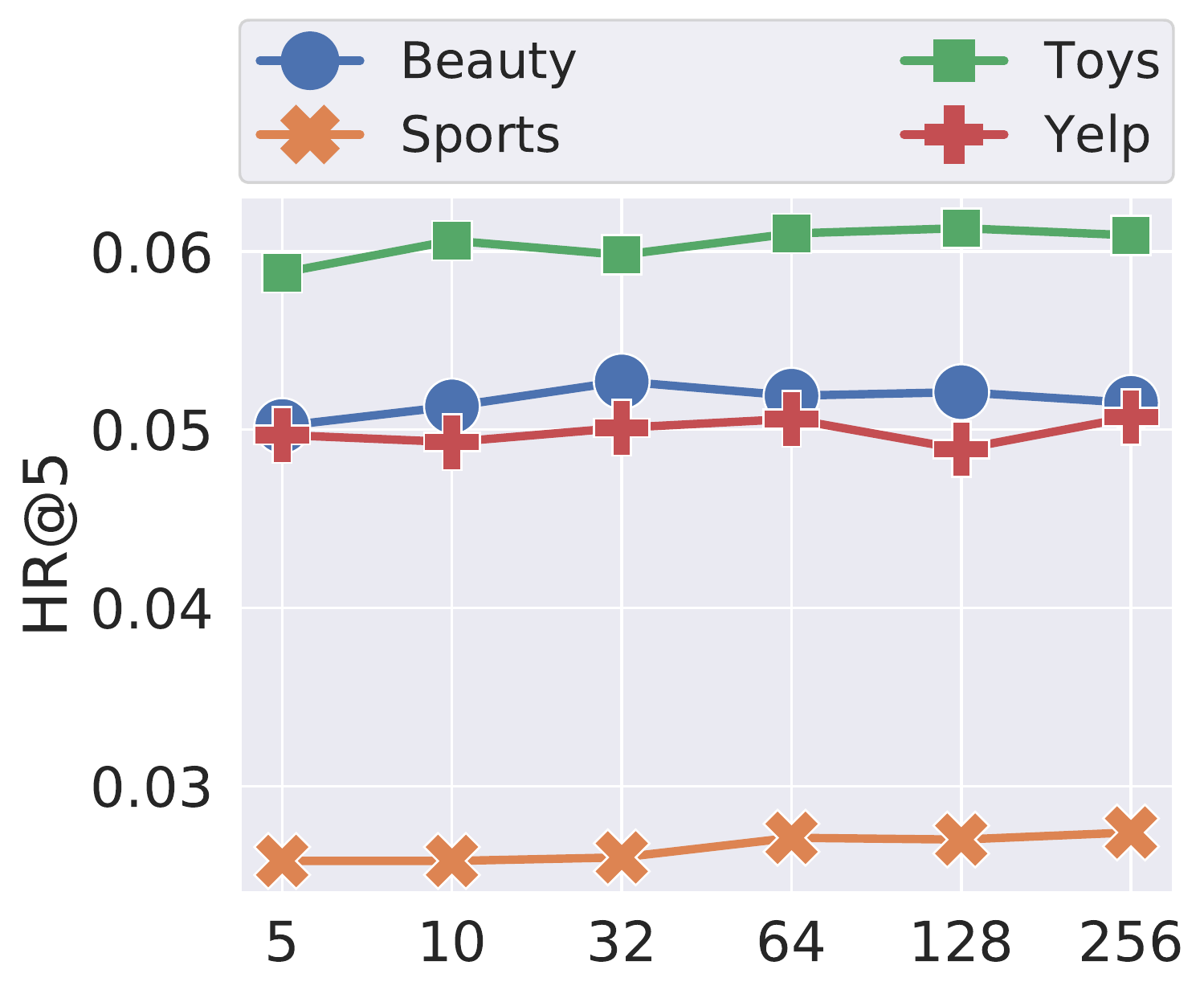}
    }
    \subfigure[NDCG@5.]{
    \includegraphics[width=0.465\linewidth]{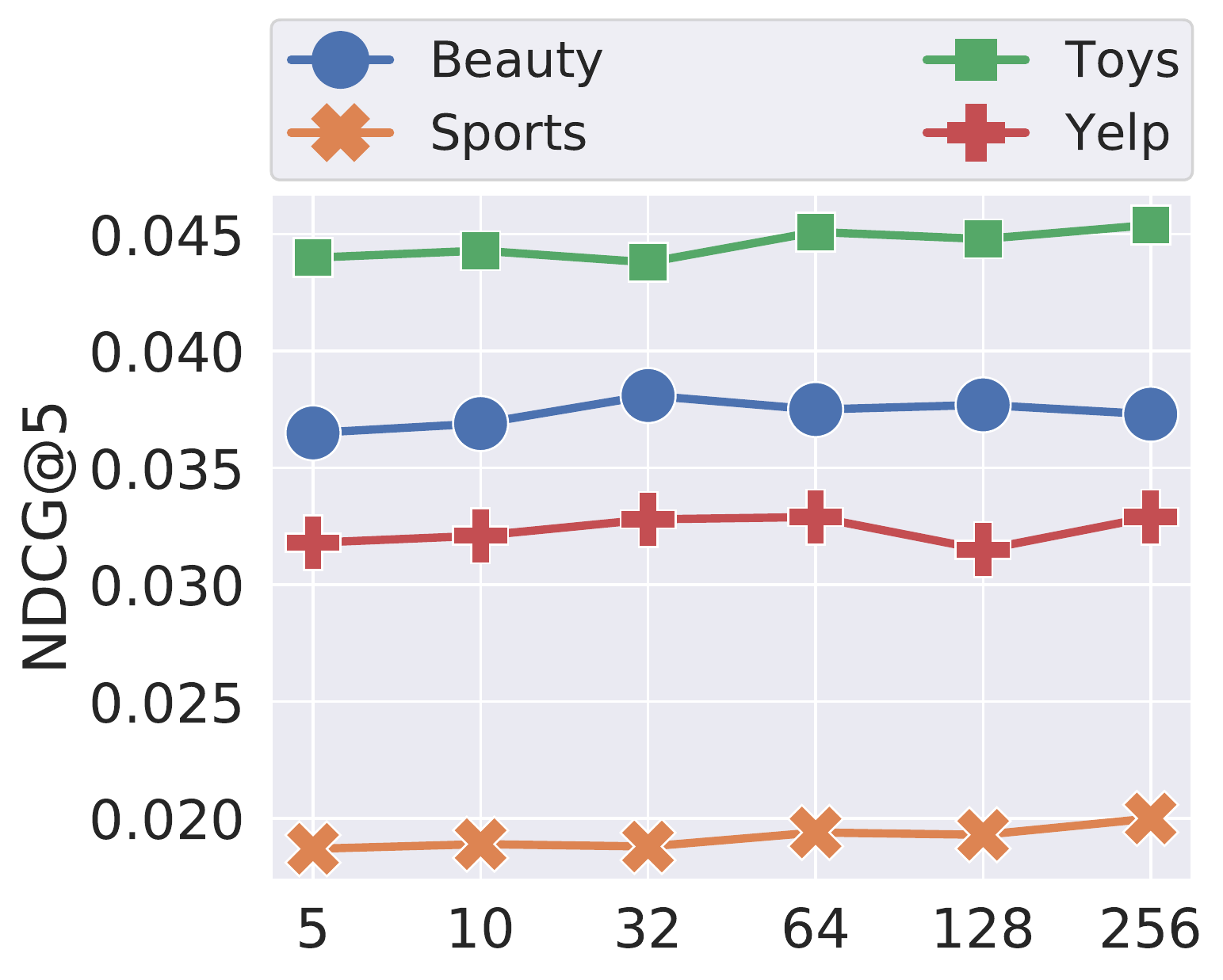}
    }
    \caption{Performance of different numbers of memory slot. The parameter $b$ is chosen from $\{5,10,32,64,128,256\}$.}
\label{fig:mem-param}
\end{figure}

\subsubsection{Impact of $b$}
In this experiment, the impact of the number of memory slot, $b$ in Eq.~(\ref{eq:mem}) is evaluated. We choose $b$ from the set $\{5,10,32,64,128,256\}$. The result is presented in Fig.~\ref{fig:mem-param}. From the result, it can be seen that the number of memory slots is not affecting the overall performance significantly. It is worth noting that according to the visualisation result of the memory bank in Fig.~\ref{fig:mem-vis}, there are only a small number of memory slots are contributing to the model. This can explain that although the number of overall memory slots is changing vastly, the performance is still very stable.

\begin{figure}[t]
    \centering
    \subfigure[HR@5.]{
    \includegraphics[width=0.465\linewidth]{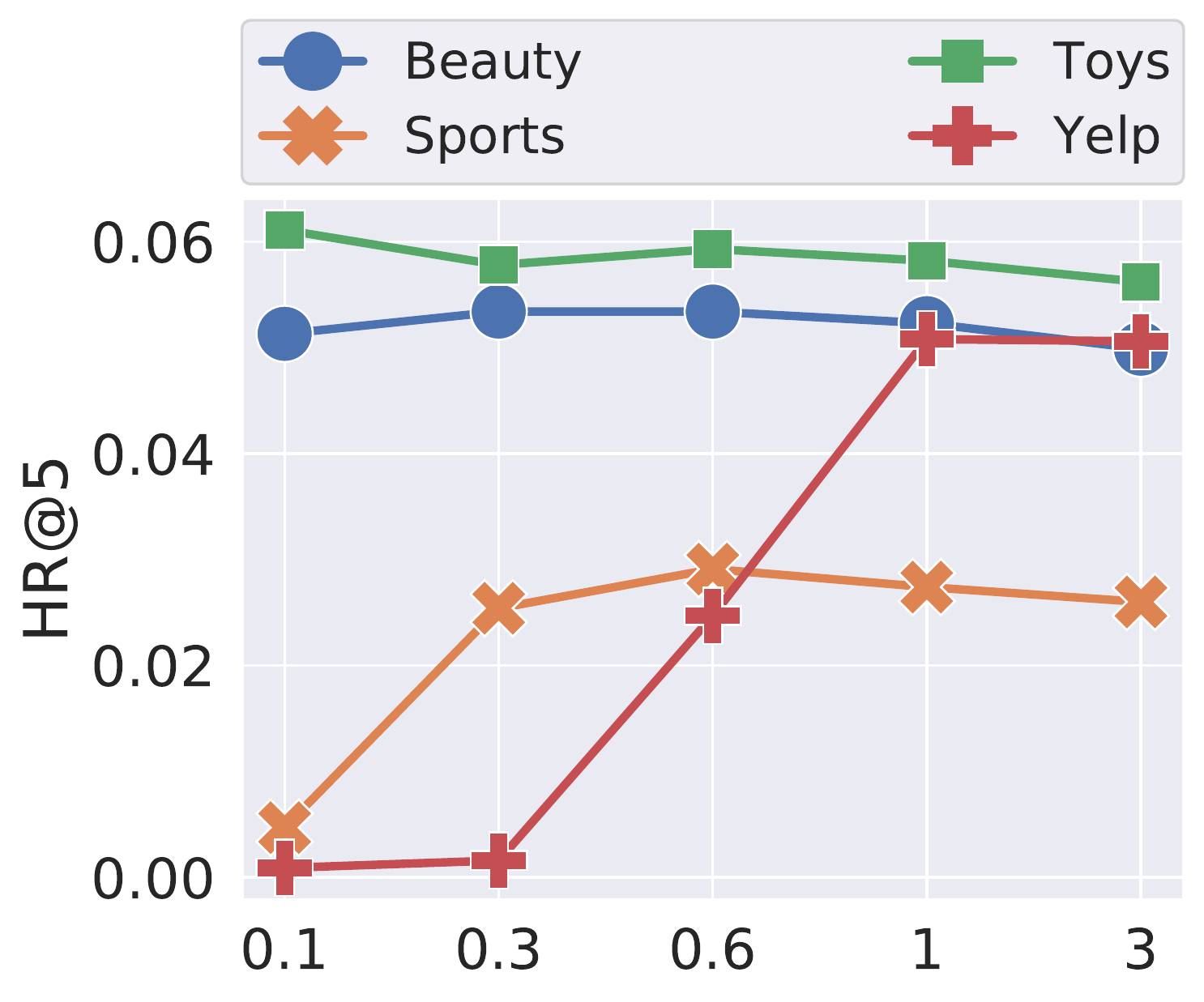}
    }
    \subfigure[NDCG@5.]{
    \includegraphics[width=0.465\linewidth]{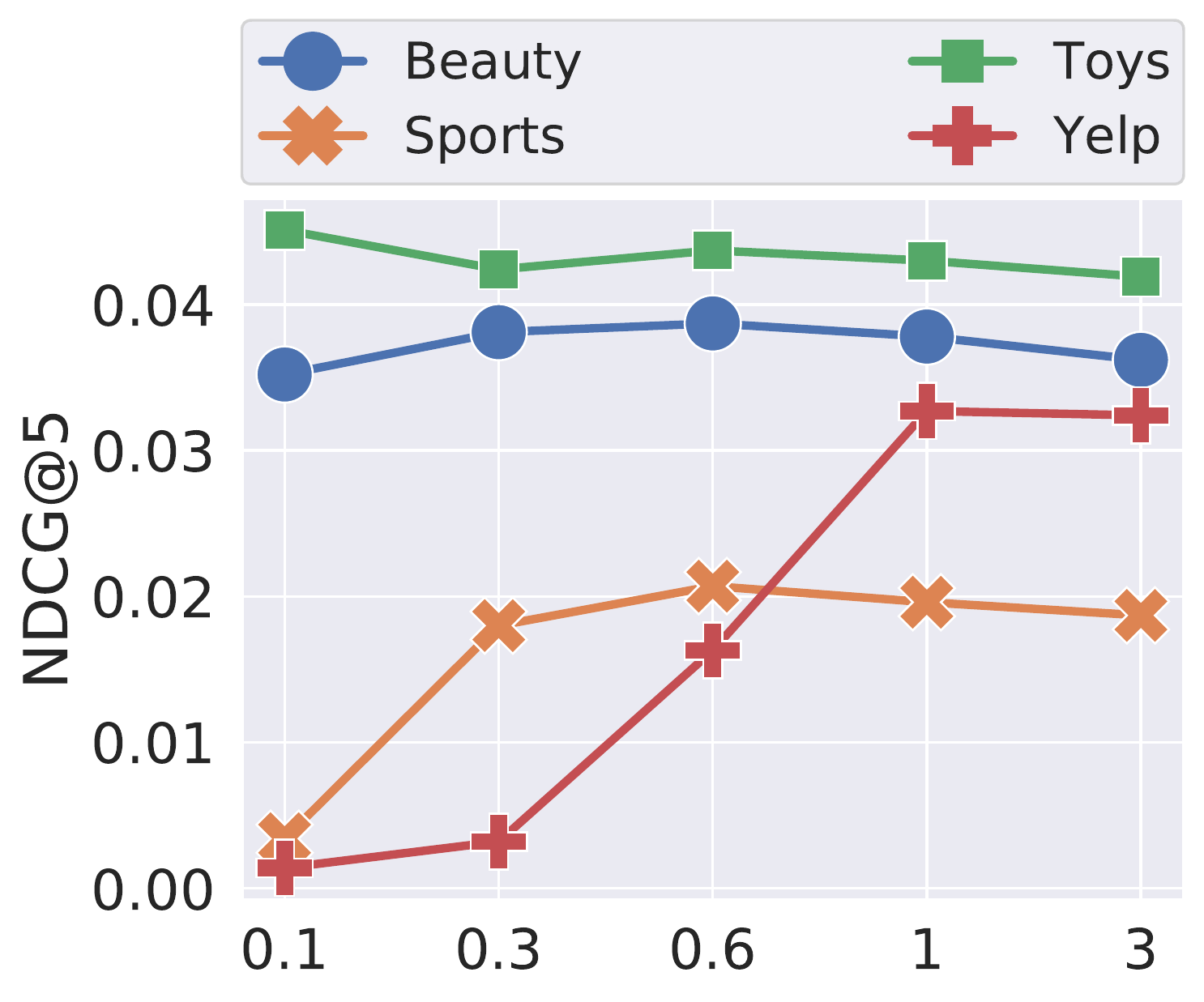}
    }
    \caption{Performance of different temperature parameter $\tau$, which is chosen from $\{0.1,0.3,0.6,1,3\}$.}
\label{fig:tau-param}
\end{figure}

\subsubsection{Impact of $\tau$}
In this experiment, the impact of the temperature parameter $\tau$ is evaluated, which is chosen from the set $\{0.1,0.3,0.6,1,3\}$. According to the experimental results in Fig.~\ref{fig:tau-param}, $\tau$ needs to set in a range to achieve a reasonable performance. For example for Sports and Yelp datasets, when $\tau$ is too small, the prediction logits become closer to a deterministic distribution, which could not provide sufficient training signals to train the model.

\begin{figure}[t]
    \centering
    \subfigure[HR@5.]{
    \includegraphics[width=0.465\linewidth]{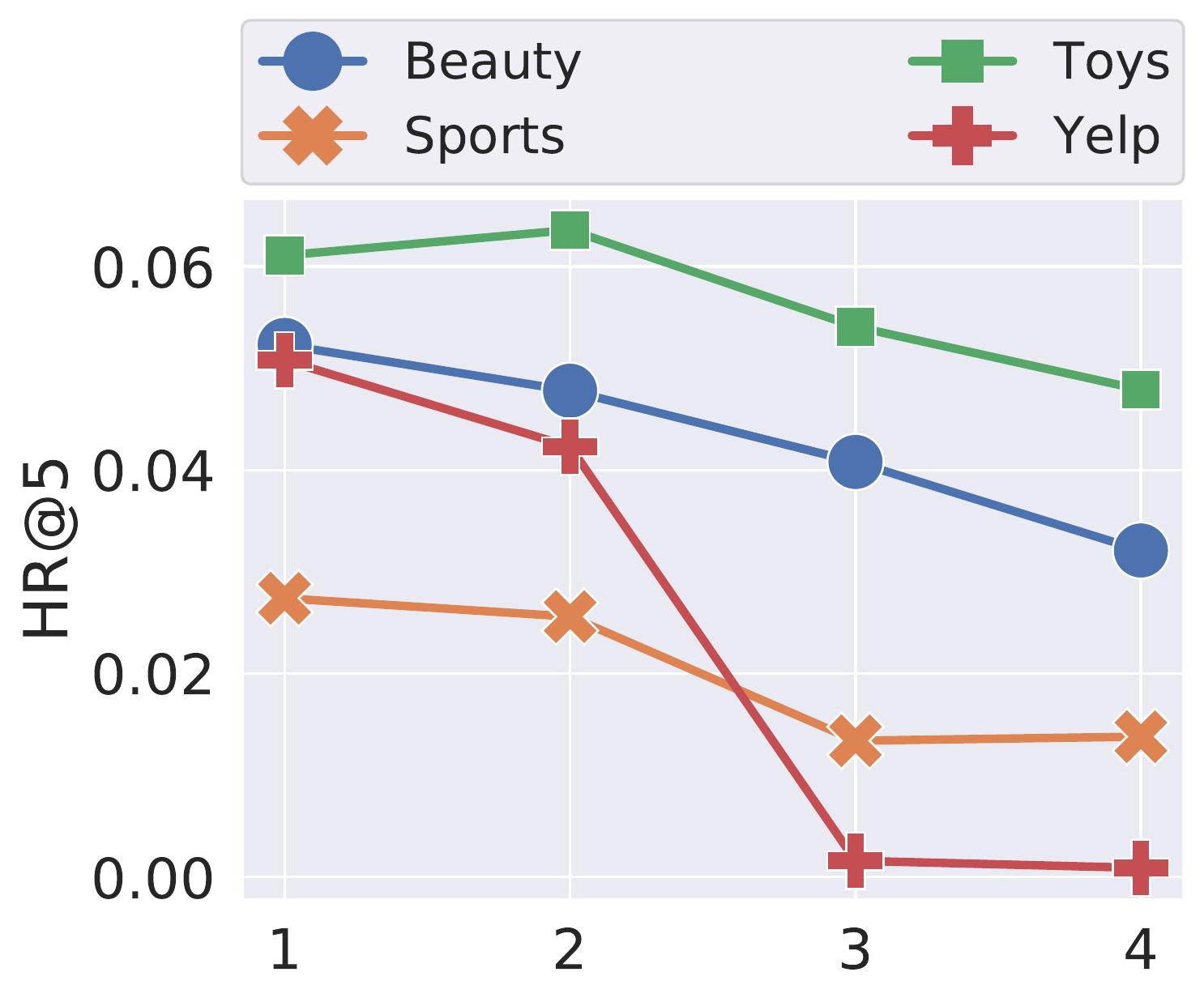}
    }
    \subfigure[NDCG@5.]{
    \includegraphics[width=0.465\linewidth]{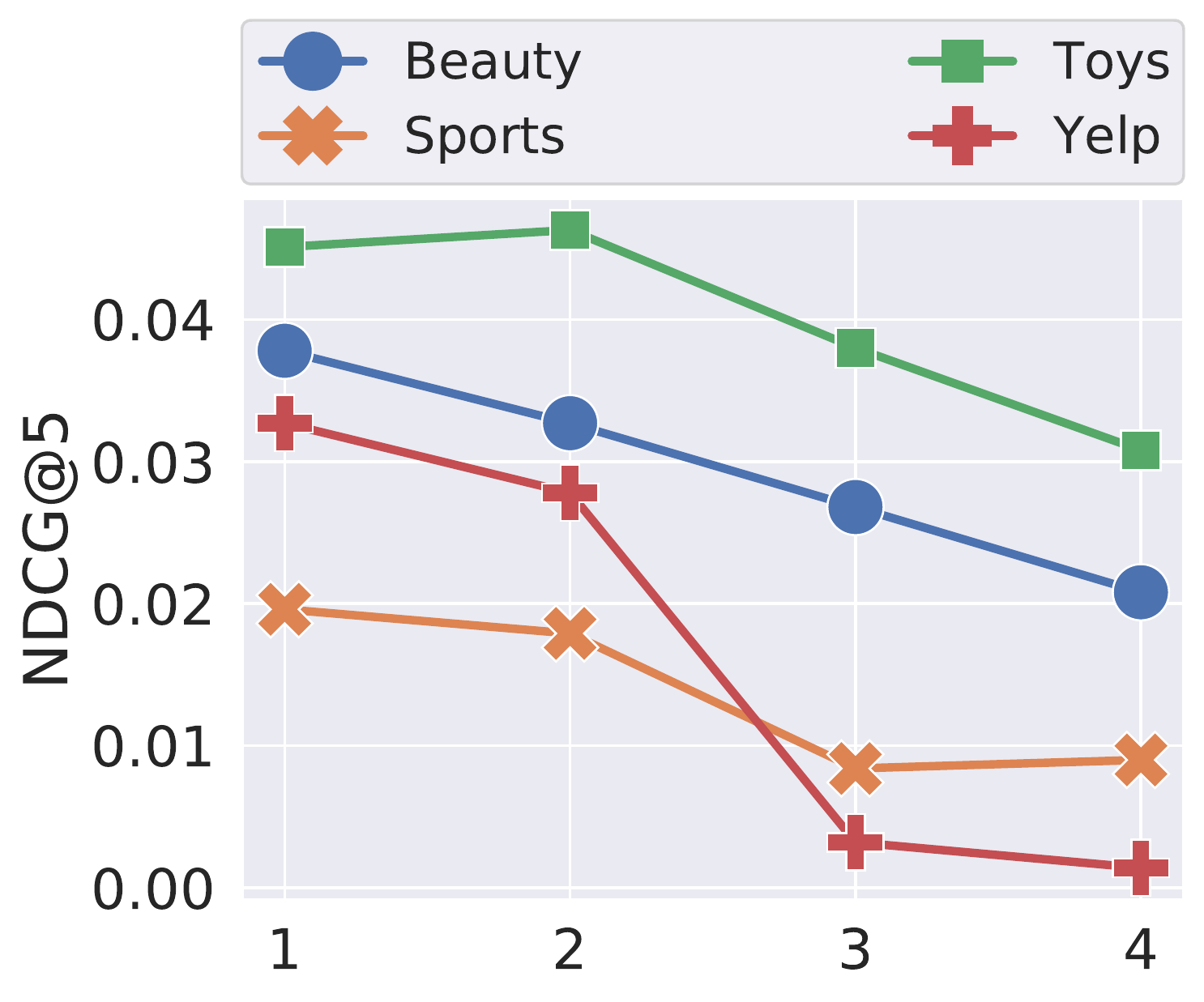}
    }
    \caption{Performance of different numbers of prediction step. The number of steps is chosen from $\{1,2,3,4\}$.}
\label{fig:pred-param}
\end{figure}

\subsubsection{Impact of number of prediction step}
In this experiment, the impact of the number of the prediction step is evaluated. This parameter controls how many steps the auto-regressive prediction $g_\text{ar}$ rolls out and how much of the futuristic information will be included into the training. From the result in Fig.~\ref{fig:pred-param}, except for the Toys dataset, MMInfoRec cannot gain useful information from the futuristic steps on all other datasets. For the Toys dataset, MMInfoRec achieves the highest performance with a two-step prediction.

\section{Conclusion}
\label{sec:conc}
In this paper, the MMInfoRec model is proposed with a memory augmented CPC training scheme and a novel MINCE loss for effective training. Specifically, the memory module is designed to provide a flexible and general representation of the sequence to provide a comprehensive long-term preference. The MINCE introduces a multi-instance noise contrastive estimation objective by using the semantically similar samples under different Dropout functions with the same item encoder. In the experiment, we successfully demonstrate that the MMInfoRec achieves the best performance compared with state-of-the-art baselines by a large margin.

\section{Acknowledgments}
The work was supported by Australian Research Council (CE200100025, DP190102353, DP190101985, FT210100624).

\bibliographystyle{IEEEtranS}
\bibliography{IEEEabrv,icdm}

\begin{thebibliography}{10}
\providecommand{\url}[1]{#1}
\csname url@samestyle\endcsname
\providecommand{\newblock}{\relax}
\providecommand{\bibinfo}[2]{#2}
\providecommand{\BIBentrySTDinterwordspacing}{\spaceskip=0pt\relax}
\providecommand{\BIBentryALTinterwordstretchfactor}{4}
\providecommand{\BIBentryALTinterwordspacing}{\spaceskip=\fontdimen2\font plus
\BIBentryALTinterwordstretchfactor\fontdimen3\font minus
  \fontdimen4\font\relax}
\providecommand{\BIBforeignlanguage}[2]{{%
\expandafter\ifx\csname l@#1\endcsname\relax
\typeout{** WARNING: IEEEtranS.bst: No hyphenation pattern has been}%
\typeout{** loaded for the language `#1'. Using the pattern for}%
\typeout{** the default language instead.}%
\else
\language=\csname l@#1\endcsname
\fi
#2}}
\providecommand{\BIBdecl}{\relax}
\BIBdecl

\bibitem{simclr}
T.~Chen, S.~Kornblith, M.~Norouzi, and G.~E. Hinton, ``A simple framework for
  contrastive learning of visual representations,'' in \emph{ICML}, 2020.

\bibitem{safm}
T.~Chen, H.~Yin, Q.~V.~H. Nguyen, W.~Peng, X.~Li, and X.~Zhou, ``Sequence-aware
  factorization machines for temporal predictive analytics,'' in \emph{ICDE},
  2020.

\bibitem{gru}
J.~Chung, C.~Gulcehre, K.~Cho, and Y.~Bengio, ``Empirical evaluation of gated
  recurrent neural networks on sequence modeling,'' in \emph{NIPS}, 2014.

\bibitem{cmn}
T.~Ebesu, B.~Shen, and Y.~Fang, ``Collaborative memory network for
  recommendation systems,'' in \emph{SIGIR}, 2018.

\bibitem{stan}
D.~Garg, P.~Gupta, P.~Malhotra, L.~Vig, and G.~Shroff, ``Sequence and time
  aware neighborhood for session-based recommendations: {STAN},'' in
  \emph{SIGIR}, 2019.

\bibitem{ntm}
A.~Graves, G.~Wayne, and I.~Danihelka, ``Neural turing machines,'' \emph{CoRR},
  vol. abs/1410.5401, 2014.

\bibitem{nce}
M.~Gutmann and A.~Hyv{\"{a}}rinen, ``Noise-contrastive estimation of
  unnormalized statistical models, with applications to natural image
  statistics,'' \emph{J. Mach. Learn. Res.}, 2012.

\bibitem{memdpc}
T.~Han, W.~Xie, and A.~Zisserman, ``Memory-augmented dense predictive coding
  for video representation learning,'' in \emph{ECCV}, 2020.

\bibitem{coclr}
------, ``Self-supervised co-training for video representation learning,'' in
  \emph{NeurIPS}, 2020.

\bibitem{moco}
K.~He, H.~Fan, Y.~Wu, S.~Xie, and R.~B. Girshick, ``Momentum contrast for
  unsupervised visual representation learning,'' in \emph{CVPR}, 2020.

\bibitem{ecpc}
O.~J. H{\'{e}}naff, A.~Srinivas, J.~D. Fauw, A.~Razavi, C.~Doersch, S.~M.~A.
  Eslami, and A.~van~den Oord, ``Data-efficient image recognition with
  contrastive predictive coding,'' \emph{CoRR}, vol. abs/1905.09272, 2019.

\bibitem{gru4rec}
B.~Hidasi, A.~Karatzoglou, L.~Baltrunas, and D.~Tikk, ``Session-based
  recommendations with recurrent neural networks,'' in \emph{ICLR}, 2016.

\bibitem{gru4recf}
B.~Hidasi, M.~Quadrana, A.~Karatzoglou, and D.~Tikk, ``Parallel recurrent
  neural network architectures for feature-rich session-based
  recommendations,'' in \emph{RecSys}, 2016.

\bibitem{deepinfomax}
R.~D. Hjelm, A.~Fedorov, S.~Lavoie{-}Marchildon, K.~Grewal, P.~Bachman,
  A.~Trischler, and Y.~Bengio, ``Learning deep representations by mutual
  information estimation and maximization,'' in \emph{ICLR}, 2019.

\bibitem{sasrec}
W.~Kang and J.~J. McAuley, ``Self-attentive sequential recommendation,'' in
  \emph{ICDM}, 2018.

\bibitem{scl}
P.~Khosla, P.~Teterwak, C.~Wang, A.~Sarna, Y.~Tian, P.~Isola, A.~Maschinot,
  C.~Liu, and D.~Krishnan, ``Supervised contrastive learning,'' in
  \emph{NeurIPS}, 2020.

\bibitem{adam}
D.~P. Kingma and J.~Ba, ``Adam: {A} method for stochastic optimization,'' in
  \emph{ICLR}, 2015.

\bibitem{infoword}
L.~Kong, C.~de~Masson~d'Autume, L.~Yu, W.~Ling, Z.~Dai, and D.~Yogatama, ``A
  mutual information maximization perspective of language representation
  learning,'' in \emph{ICLR}, 2020.

\bibitem{metric}
W.~Krichene and S.~Rendle, ``On sampled metrics for item recommendation,'' in
  \emph{SIGKDD}, 2020.

\bibitem{seq2graph}
Y.~Li, T.~Chen, Y.~Luo, H.~Yin, and Z.~Huang, ``Discovering collaborative
  signals for next {POI} recommendation with iterative seq2graph
  augmentation,'' in \emph{IJCAI}, 2021.

\bibitem{lightweight}
Y.~Li, T.~Chen, P.~Zhang, and H.~Yin, ``Lightweight self-attentive sequential
  recommendation,'' \emph{CoRR}, vol. abs/2108.11333, 2021.

\bibitem{hgn}
C.~Ma, P.~Kang, and X.~Liu, ``Hierarchical gating networks for sequential
  recommendation,'' in \emph{SIGKDD}, 2019.

\bibitem{magnn}
C.~Ma, L.~Ma, Y.~Zhang, J.~Sun, X.~Liu, and M.~Coates, ``Memory augmented graph
  neural networks for sequential recommendation,'' in \emph{AAAI}, 2020.

\bibitem{s2s}
J.~Ma, C.~Zhou, H.~Yang, P.~Cui, X.~Wang, and W.~Zhu, ``Disentangled
  self-supervision in sequential recommenders,'' in \emph{SIGKDD}, 2020.

\bibitem{amazon}
J.~J. McAuley, C.~Targett, Q.~Shi, and A.~van~den Hengel, ``Image-based
  recommendations on styles and substitutes,'' in \emph{SIGIR}, 2015.

\bibitem{milnce}
A.~Miech, J.~Alayrac, L.~Smaira, I.~Laptev, J.~Sivic, and A.~Zisserman,
  ``End-to-end learning of visual representations from uncurated instructional
  videos,'' in \emph{CVPR}, 2020.

\bibitem{nce2}
A.~Mnih and K.~Kavukcuoglu, ``Learning word embeddings efficiently with
  noise-contrastive estimation,'' in \emph{NIPS}, 2013.

\bibitem{vbmi}
B.~Poole, S.~Ozair, A.~van~den Oord, A.~Alemi, and G.~Tucker, ``On variational
  bounds of mutual information,'' in \emph{ICML}, 2019.

\bibitem{posrec}
R.~Qiu, Z.~Huang, T.~Chen, and H.~Yin, ``Exploiting positional information for
  session-based recommendation,'' \emph{CoRR}, vol. abs/2107.00846, 2021.

\bibitem{fgnnj}
R.~Qiu, Z.~Huang, J.~Li, and H.~Yin, ``Exploiting cross-session information for
  session-based recommendation with graph neural networks,'' \emph{{ACM} Trans.
  Inf. Syst.}, vol.~38, no.~3, pp. 22:1--22:23, 2020.

\bibitem{fgnn}
R.~Qiu, J.~Li, Z.~Huang, and H.~Yin, ``Rethinking the item order in
  session-based recommendation with graph neural networks,'' in \emph{CIKM},
  2019.

\bibitem{causalrec}
R.~Qiu, S.~Wang, Z.~Chen, H.~Yin, and Z.~Huang, ``Causalrec: Causal inference
  for visual debiasing in visually-aware recommendation,'' \emph{CoRR}, vol.
  abs/2107.02390, 2021.

\bibitem{gag}
R.~Qiu, H.~Yin, Z.~Huang, and T.~Chen, ``{GAG:} global attributed graph neural
  network for streaming session-based recommendation,'' in \emph{SIGIR}, 2020.

\bibitem{bprmf}
S.~Rendle, C.~Freudenthaler, Z.~Gantner, and L.~Schmidt{-}Thieme, ``{BPR:}
  bayesian personalized ranking from implicit feedback,'' in \emph{UAI}, 2009.

\bibitem{autoint}
W.~Song, C.~Shi, Z.~Xiao, Z.~Duan, Y.~Xu, M.~Zhang, and J.~Tang, ``Autoint:
  Automatic feature interaction learning via self-attentive neural networks,''
  in \emph{CIKM}, 2019.

\bibitem{dropout}
N.~Srivastava, G.~Hinton, A.~Krizhevsky, I.~Sutskever, and R.~Salakhutdinov,
  ``Dropout: A simple way to prevent neural networks from overfitting,''
  \emph{Journal of Machine Learning Research}, 2014.

\bibitem{bert4rec}
F.~Sun, J.~Liu, J.~Wu, C.~Pei, X.~Lin, W.~Ou, and P.~Jiang, ``Bert4rec:
  Sequential recommendation with bidirectional encoder representations from
  transformer,'' in \emph{CIKM}, 2019.

\bibitem{seq2seq}
I.~Sutskever, O.~Vinyals, and Q.~V. Le, ``Sequence to sequence learning with
  neural networks,'' in \emph{NIPS}, 2014.

\bibitem{dman}
Q.~Tan, J.~Zhang, N.~Liu, X.~Huang, H.~Yang, J.~Zhou, and X.~Hu, ``Dynamic
  memory based attention network for sequential recommendation,'' in
  \emph{AAAI}, 2021.

\bibitem{caser}
J.~Tang and K.~Wang, ``Personalized top-n sequential recommendation via
  convolutional sequence embedding,'' in \emph{WSDM}, 2018.

\bibitem{cpc}
A.~van~den Oord, Y.~Li, and O.~Vinyals, ``Representation learning with
  contrastive predictive coding,'' \emph{CoRR}, vol. abs/1807.03748, 2018.

\bibitem{attention}
A.~Vaswani, N.~Shazeer, N.~Parmar, J.~Uszkoreit, L.~Jones, A.~N. Gomez,
  L.~Kaiser, and I.~Polosukhin, ``Attention is all you need,'' in \emph{NIPS},
  2017.

\bibitem{nmrn}
Q.~Wang, H.~Yin, Z.~Hu, D.~Lian, H.~Wang, and Z.~Huang, ``Neural memory
  streaming recommender networks with adversarial training,'' in \emph{SIGKDD},
  2018.

\bibitem{ldsdg}
Z.~Wang, Y.~Luo, R.~Qiu, Z.~Huang, and M.~Baktashmotlagh, ``Learning to
  diversify for single domain generalization,'' \emph{CoRR}, vol.
  abs/2108.11726, 2021.

\bibitem{matn}
L.~Xia, C.~Huang, Y.~Xu, P.~Dai, B.~Zhang, and L.~Bo, ``Multiplex behavioral
  relation learning for recommendation via memory augmented transformer
  network,'' in \emph{SIGIR}, 2020.

\bibitem{hypergraph}
X.~Xia, H.~Yin, J.~Yu, Q.~Wang, L.~Cui, and X.~Zhang, ``Self-supervised
  hypergraph convolutional networks for session-based recommendation,'' in
  \emph{AAAI}, 2021.

\bibitem{xie2020contrastive}
X.~Xie, F.~Sun, Z.~Liu, S.~Wu, J.~Gao, B.~Ding, and B.~Cui, ``Contrastive
  learning for sequential recommendation,'' \emph{CoRR}, vol. abs/2010.14395,
  2021.

\bibitem{channel}
J.~Yu, H.~Yin, J.~Li, Q.~Wang, N.~Q.~V. Hung, and X.~Zhang, ``Self-supervised
  multi-channel hypergraph convolutional network for social recommendation,''
  in \emph{WWW}, 2021.

\bibitem{fdsa}
T.~Zhang, P.~Zhao, Y.~Liu, V.~S. Sheng, J.~Xu, D.~Wang, G.~Liu, and X.~Zhou,
  ``Feature-level deeper self-attention network for sequential
  recommendation,'' in \emph{IJCAI}, 2019.

\bibitem{zhou2020contrastive}
C.~Zhou, J.~Ma, J.~Zhang, J.~Zhou, and H.~Yang, ``Contrastive learning for
  debiased candidate generation in large-scale recommender systems,'' in
  \emph{SIGKDD}, 2021.

\bibitem{s3rec}
K.~Zhou, H.~Wang, W.~X. Zhao, Y.~Zhu, S.~Wang, F.~Zhang, Z.~Wang, and J.~Wen,
  ``S{\^{}}3-rec: Self-supervised learning for sequential recommendation with
  mutual information maximization,'' in \emph{CIKM}, 2020.

\end{thebibliography}

\end{document}